\documentstyle[12pt,floatfig,epsfig,citequo]{article}
\epsfverbosetrue
\def\EPJ{Eur. Phys. J.}

\def\PL{Phys. Lett.}
\def\PR{Phys. Rev.}

\def\ZP{Z. Phys.}

\begin{document}

\begin{center}
{\large EUROPEAN LABORATORY FOR PARTICLE PHYSICS}
\end{center}

\vspace{0.5cm}
\begin{flushright}CERN-EP/99-025\\
22 February 1999\\
\end{flushright}

\begin{center}
\vspace{0.5cm}
{\huge 
\bf
One-prong $\tau$ decays with kaons}
\end{center}

\begin{center}
\vspace{1cm}
\bf {The ALEPH Collaboration}
\end{center}

\vspace{0.5cm}
\begin{abstract}
One-prong $\tau$ decays into final states involving kaons are
studied with about 161k $\tau^+\tau^-$ events collected by the ALEPH 
detector from 1991 to 1995. Charged kaons are identified by dE/dx measurement,
while $K^0_L$'s are detected through their interaction in calorimeters.
Branching ratios are measured for the inclusive mode,
$B(\tau^-\rightarrow K^-X\nu_\tau)=(1.52 \pm 0.04\pm0.04)\%$, where $X$ can
be any system of neutral particles, and for the exclusive modes
\begin{center}
\begin{tabular}{rcl}
$B(\tau^-\to K^-\nu_\tau)$ &=& $(6.96\pm0.25\pm0.14)\times 10^{-3}$,\\
$B(\tau^-\to K^-\pi^0\nu_\tau)$ &=& $(4.44\pm0.26\pm0.24)\times 10^{-3}$,\\
$B(\tau^-\to K^-\pi^0\pi^0\nu_\tau)$ &=&
$(0.56\pm0.20\pm0.15)\times 10^{-3}$,\\
$B(\tau^-\to K^-\pi^0\pi^0\pi^0\nu_\tau)$ &=&
$(0.37\pm0.21\pm0.11)\times 10^{-3}$,\\
$B(\tau^-\to K^-K^{0}\nu_\tau)$ &=& $(1.62\pm0.21\pm0.11)\times 10^{-3}$,\\
$B(\tau^-\to K^{-}K^{0}\pi^0\nu_\tau)$ &
=& $(1.43\pm 0.25\pm0.21)\times 10^{-3}$,\\
$B(\tau^-\to \overline{K^0}
\pi^{-}\nu_\tau)$ &=& $(9.28\pm 0.45\pm0.34)\times 10^{-3}$,\\
$B(\tau^-\to \overline{K^0}
\pi^{-}\pi^0\nu_\tau)$ &=& $(3.47\pm0.53\pm0.37)\times10^{-3}$.
\end{tabular}
\end{center}
Upper limits for
$B(\tau^-\to \overline{K^0}\pi^{-}\pi^0\pi^0\nu_\tau)$ and
$B(\tau^-\to K^{-}K^{0}\pi^0\pi^0\nu_\tau)$ are also obtained.
Mass spectra in the final states are
addressed in order to study the relevant dynamics.
\end{abstract}
\vspace{1cm}
\centerline{\it (To be submitted to European Physical Journal C) }
\pagestyle{empty}
\newpage
\small
%
\newlength{\saveparskip}
\newlength{\savetextheight}
\newlength{\savetopmargin}
\newlength{\savetextwidth}
\newlength{\saveoddsidemargin}
\newlength{\savetopsep}
\setlength{\saveparskip}{\parskip}
\setlength{\savetextheight}{\textheight}
\setlength{\savetopmargin}{\topmargin}
\setlength{\savetextwidth}{\textwidth}
\setlength{\saveoddsidemargin}{\oddsidemargin}
\setlength{\savetopsep}{\topsep}
%
%
\setlength{\parskip}{0.0cm}
\setlength{\textheight}{25.0cm}
\setlength{\topmargin}{-1.5cm}
\setlength{\textwidth}{16 cm}
\setlength{\oddsidemargin}{-0.0cm}
\setlength{\topsep}{1mm}
\pretolerance=10000
\centerline{\large\bf The ALEPH Collaboration}
\footnotesize
\vspace{0.5cm}
{\raggedbottom
\begin{sloppypar}
\samepage\noindent
R.~Barate,
D.~Decamp,
P.~Ghez,
C.~Goy,
\mbox{J.-P.~Lees},
E.~Merle,
\mbox{M.-N.~Minard},
B.~Pietrzyk
\nopagebreak
\begin{center}
\parbox{15.5cm}{\sl\samepage
Laboratoire de Physique des Particules (LAPP), IN$^{2}$P$^{3}$-CNRS,
F-74019 Annecy-le-Vieux Cedex, France}
\end{center}\end{sloppypar}
\vspace{2mm}
\begin{sloppypar}
\noindent
R.~Alemany,
M.P.~Casado,
M.~Chmeissani,
J.M.~Crespo,
E.~Fernandez,
\mbox{M.~Fernandez-Bosman},
Ll.~Garrido,$^{15}$
E.~Graug\`{e}s,
A.~Juste,
M.~Martinez,
G.~Merino,
R.~Miquel,
Ll.M.~Mir,
A.~Pacheco,
I.C.~Park,
I.~Riu
\nopagebreak
\begin{center}
\parbox{15.5cm}{\sl\samepage
Institut de F\'{i}sica d'Altes Energies, Universitat Aut\`{o}noma
de Barcelona, E-08193 Bellaterra (Barcelona), Spain$^{7}$}
\end{center}\end{sloppypar}
\vspace{2mm}
\begin{sloppypar}
\noindent
A.~Colaleo,
D.~Creanza,
M.~de~Palma,
G.~Gelao,
G.~Iaselli,
G.~Maggi,
M.~Maggi,
S.~Nuzzo,
A.~Ranieri,
G.~Raso,
F.~Ruggieri,
G.~Selvaggi,
L.~Silvestris,
P.~Tempesta,
A.~Tricomi,$^{3}$
G.~Zito
\nopagebreak
\begin{center}
\parbox{15.5cm}{\sl\samepage
Dipartimento di Fisica, INFN Sezione di Bari, I-70126
Bari, Italy}
\end{center}\end{sloppypar}
\vspace{2mm}
\begin{sloppypar}
\noindent
X.~Huang,
J.~Lin,
Q. Ouyang,
T.~Wang,
Y.~Xie,
R.~Xu,
S.~Xue,
J.~Zhang,
L.~Zhang,
W.~Zhao
\nopagebreak
\begin{center}
\parbox{15.5cm}{\sl\samepage
Institute of High-Energy Physics, Academia Sinica, Beijing, The People's
Republic of China$^{8}$}
\end{center}\end{sloppypar}
\vspace{2mm}
\begin{sloppypar}
\noindent
D.~Abbaneo,
U.~Becker,$^{19}$
G.~Boix,$^{6}$
M.~Cattaneo,
V.~Ciulli,
G.~Dissertori,
H.~Drevermann,
R.W.~Forty,
M.~Frank,
A.W. Halley,
J.B.~Hansen,
J.~Harvey,
P.~Janot,
B.~Jost,
I.~Lehraus,
O.~Leroy,
P.~Mato,
A.~Minten,
A.~Moutoussi,
F.~Ranjard,
L.~Rolandi,
D.~Rousseau,
D.~Schlatter,
M.~Schmitt,$^{20}$
O.~Schneider,$^{23}$
W.~Tejessy,
F.~Teubert,
I.R.~Tomalin,
E.~Tournefier,
A.E.~Wright
\nopagebreak
\begin{center}
\parbox{15.5cm}{\sl\samepage
European Laboratory for Particle Physics (CERN), CH-1211 Geneva 23,
Switzerland}
\end{center}\end{sloppypar}
\vspace{2mm}
\begin{sloppypar}
\noindent
Z.~Ajaltouni,
F.~Badaud,
G.~Chazelle,
O.~Deschamps,
A.~Falvard,
C.~Ferdi,
P.~Gay,
C.~Guicheney,
P.~Henrard,
J.~Jousset,
B.~Michel,
S.~Monteil,
\mbox{J-C.~Montret},
D.~Pallin,
P.~Perret,
F.~Podlyski
\nopagebreak
\begin{center}
\parbox{15.5cm}{\sl\samepage
Laboratoire de Physique Corpusculaire, Universit\'e Blaise Pascal,
IN$^{2}$P$^{3}$-CNRS, Clermont-Ferrand, F-63177 Aubi\`{e}re, France}
\end{center}\end{sloppypar}
\vspace{2mm}
\begin{sloppypar}
\noindent
J.D.~Hansen,
J.R.~Hansen,
P.H.~Hansen,
B.S.~Nilsson,
B.~Rensch,
A.~W\"a\"an\"anen
\begin{center}
\parbox{15.5cm}{\sl\samepage
Niels Bohr Institute, DK-2100 Copenhagen, Denmark$^{9}$}
\end{center}\end{sloppypar}
\vspace{2mm}
\begin{sloppypar}
\noindent
G.~Daskalakis,
A.~Kyriakis,
C.~Markou,
E.~Simopoulou,
I.~Siotis,
A.~Vayaki
\nopagebreak
\begin{center}
\parbox{15.5cm}{\sl\samepage
Nuclear Research Center Demokritos (NRCD), GR-15310 Attiki, Greece}
\end{center}\end{sloppypar}
\vspace{2mm}
\begin{sloppypar}
\noindent
A.~Blondel,
G.~Bonneaud,
\mbox{J.-C.~Brient},
A.~Roug\'{e},
M.~Rumpf,
M.~Swynghedauw,
M.~Verderi,
H.~Videau
\nopagebreak
\begin{center}
\parbox{15.5cm}{\sl\samepage
Laboratoire de Physique Nucl\'eaire et des Hautes Energies, Ecole
Polytechnique, IN$^{2}$P$^{3}$-CNRS, \mbox{F-91128} Palaiseau Cedex, France}
\end{center}\end{sloppypar}
\vspace{2mm}
\begin{sloppypar}
\noindent
E.~Focardi,
G.~Parrini,
K.~Zachariadou
\nopagebreak
\begin{center}
\parbox{15.5cm}{\sl\samepage
Dipartimento di Fisica, Universit\`a di Firenze, INFN Sezione di Firenze,
I-50125 Firenze, Italy}
\end{center}\end{sloppypar}
\vspace{2mm}
\begin{sloppypar}
\noindent
R.~Cavanaugh,
M.~Corden,
C.~Georgiopoulos
\nopagebreak
\begin{center}
\parbox{15.5cm}{\sl\samepage
Supercomputer Computations Research Institute,
Florida State University,
Tallahassee, FL 32306-4052, USA $^{13,14}$}
\end{center}\end{sloppypar}
\vspace{2mm}
\begin{sloppypar}
\noindent
A.~Antonelli,
G.~Bencivenni,
G.~Bologna,$^{4}$
F.~Bossi,
P.~Campana,
G.~Capon,
F.~Cerutti,
V.~Chiarella,
P.~Laurelli,
G.~Mannocchi,$^{5}$
F.~Murtas,
G.P.~Murtas,
L.~Passalacqua,
\mbox{M.~Pepe-Altarelli}$^{1}$
\nopagebreak
\begin{center}
\parbox{15.5cm}{\sl\samepage
Laboratori Nazionali dell'INFN (LNF-INFN), I-00044 Frascati, Italy}
\end{center}\end{sloppypar}
\vspace{2mm}
\begin{sloppypar}
\noindent
L.~Curtis,
J.G.~Lynch,
P.~Negus,
V.~O'Shea,
C.~Raine,
\mbox{P.~Teixeira-Dias},
A.S.~Thompson
\nopagebreak
\begin{center}
\parbox{15.5cm}{\sl\samepage
Department of Physics and Astronomy, University of Glasgow, Glasgow G12
8QQ,United Kingdom$^{10}$}
\end{center}\end{sloppypar}
\vspace{2mm}
\begin{sloppypar}
\noindent
O.~Buchm\"uller,
S.~Dhamotharan,
C.~Geweniger,
P.~Hanke,
G.~Hansper,
V.~Hepp,
E.E.~Kluge,
A.~Putzer,
J.~Sommer,
K.~Tittel,
S.~Werner,$^{19}$
M.~Wunsch
\nopagebreak
\begin{center}
\parbox{15.5cm}{\sl\samepage
Institut f\"ur Hochenergiephysik, Universit\"at Heidelberg, D-69120
Heidelberg, Germany$^{16}$}
\end{center}\end{sloppypar}
\vspace{2mm}
\begin{sloppypar}
\noindent
R.~Beuselinck,
D.M.~Binnie,
W.~Cameron,
P.J.~Dornan,$^{1}$
M.~Girone,
S.~Goodsir,
E.B.~Martin,
N.~Marinelli,
J.K.~Sedgbeer,
P.~Spagnolo,
E.~Thomson,
M.D.~Williams
\nopagebreak
\begin{center}
\parbox{15.5cm}{\sl\samepage
Department of Physics, Imperial College, London SW7 2BZ,
United Kingdom$^{10}$}
\end{center}\end{sloppypar}
\vspace{2mm}
\begin{sloppypar}
\noindent
V.M.~Ghete,
P.~Girtler,
E.~Kneringer,
D.~Kuhn,
G.~Rudolph
\nopagebreak
\begin{center}
\parbox{15.5cm}{\sl\samepage
Institut f\"ur Experimentalphysik, Universit\"at Innsbruck, A-6020
Innsbruck, Austria$^{18}$}
\end{center}\end{sloppypar}
\vspace{2mm}
\begin{sloppypar}
\noindent
A.P.~Betteridge,
C.K.~Bowdery,
P.G.~Buck,
P.~Colrain,
G.~Crawford,
A.J.~Finch,
F.~Foster,
G.~Hughes,
R.W.L.~Jones,
N.A.~Robertson,
M.I.~Williams
\nopagebreak
\begin{center}
\parbox{15.5cm}{\sl\samepage
Department of Physics, University of Lancaster, Lancaster LA1 4YB,
United Kingdom$^{10}$}
\end{center}\end{sloppypar}
\vspace{2mm}
\begin{sloppypar}
\noindent
I.~Giehl,
C.~Hoffmann,
K.~Jakobs,
K.~Kleinknecht,
G.~Quast,
B.~Renk,
E.~Rohne,
\mbox{H.-G.~Sander},
P.~van~Gemmeren,
H.~Wachsmuth,
C.~Zeitnitz
\nopagebreak
\begin{center}
\parbox{15.5cm}{\sl\samepage
Institut f\"ur Physik, Universit\"at Mainz, D-55099 Mainz, Germany$^{16}$}
\end{center}\end{sloppypar}
\vspace{2mm}
\begin{sloppypar}
\noindent
J.J.~Aubert,
C.~Benchouk,
A.~Bonissent,
J.~Carr,$^{1}$
P.~Coyle,
F.~Etienne,
F.~Motsch,
P.~Payre,
M.~Talby,
M.~Thulasidas
\nopagebreak
\begin{center}
\parbox{15.5cm}{\sl\samepage
Centre de Physique des Particules, Facult\'e des Sciences de Luminy,
IN$^{2}$P$^{3}$-CNRS, F-13288 Marseille, France}
\end{center}\end{sloppypar}
\vspace{2mm}
\begin{sloppypar}
\noindent
M.~Aleppo,
M.~Antonelli,
F.~Ragusa
\nopagebreak
\begin{center}
\parbox{15.5cm}{\sl\samepage
Dipartimento di Fisica, Universit\`a di Milano e INFN Sezione di Milano,
I-20133 Milano, Italy}
\end{center}\end{sloppypar}
\vspace{2mm}
\begin{sloppypar}
\noindent
R.~Berlich,
V.~B\"uscher,
H.~Dietl,
G.~Ganis,
K.~H\"uttmann,
G.~L\"utjens,
C.~Mannert,
W.~M\"anner,
\mbox{H.-G.~Moser},
S.~Schael,
R.~Settles,
H.~Seywerd,
H.~Stenzel,
W.~Wiedenmann,
G.~Wolf
\nopagebreak
\begin{center}
\parbox{15.5cm}{\sl\samepage
Max-Planck-Institut f\"ur Physik, Werner-Heisenberg-Institut,
D-80805 M\"unchen, Germany\footnotemark[16]}
\end{center}\end{sloppypar}
\vspace{2mm}
\begin{sloppypar}
\noindent
P.~Azzurri,
J.~Boucrot,
O.~Callot,
S.~Chen,
A.~Cordier,
M.~Davier,
L.~Duflot,
\mbox{J.-F.~Grivaz},
Ph.~Heusse,
A.~H\"ocker,
A.~Jacholkowska,
D.W.~Kim,$^{12}$
F.~Le~Diberder,
J.~Lefran\c{c}ois,
\mbox{A.-M.~Lutz},
\mbox{M.-H.~Schune},
\mbox{J.-J.~Veillet},
I.~Videau,$^{1}$
D.~Zerwas
\nopagebreak
\begin{center}
\parbox{15.5cm}{\sl\samepage
Laboratoire de l'Acc\'el\'erateur Lin\'eaire, Universit\'e de Paris-Sud,
IN$^{2}$P$^{3}$-CNRS, F-91898 Orsay Cedex, France}
\end{center}\end{sloppypar}
\vspace{2mm}
\begin{sloppypar}
\noindent
G.~Bagliesi,
S.~Bettarini,
T.~Boccali,
C.~Bozzi,$^{24}$
G.~Calderini,
R.~Dell'Orso,
I.~Ferrante,
L.~Fo\`{a},
A.~Giassi,
A.~Gregorio,
F.~Ligabue,
A.~Lusiani,
P.S.~Marrocchesi,
A.~Messineo,
F.~Palla,
G.~Rizzo,
G.~Sanguinetti,
A.~Sciab\`a,
G.~Sguazzoni,
R.~Tenchini,
C.~Vannini,
A.~Venturi,
P.G.~Verdini
\samepage
\begin{center}
\parbox{15.5cm}{\sl\samepage
Dipartimento di Fisica dell'Universit\`a, INFN Sezione di Pisa,
e Scuola Normale Superiore, I-56010 Pisa, Italy}
\end{center}\end{sloppypar}
\vspace{2mm}
\begin{sloppypar}
\noindent
G.A.~Blair,
G.~Cowan,
M.G.~Green,
T.~Medcalf,
J.A.~Strong,
\mbox{J.H.~von~Wimmersperg-Toeller}
\nopagebreak
\begin{center}
\parbox{15.5cm}{\sl\samepage
Department of Physics, Royal Holloway \& Bedford New College,
University of London, Surrey TW20 OEX, United Kingdom$^{10}$}
\end{center}\end{sloppypar}
\vspace{2mm}
\begin{sloppypar}
\noindent
D.R.~Botterill,
R.W.~Clifft,
T.R.~Edgecock,
P.R.~Norton,
J.C.~Thompson
\nopagebreak
\begin{center}
\parbox{15.5cm}{\sl\samepage
Particle Physics Dept., Rutherford Appleton Laboratory,
Chilton, Didcot, Oxon OX11 OQX, United Kingdom$^{10}$}
\end{center}\end{sloppypar}
\vspace{2mm}
\begin{sloppypar}
\noindent
\mbox{B.~Bloch-Devaux},
P.~Colas,
S.~Emery,
W.~Kozanecki,
E.~Lan\c{c}on,
\mbox{M.-C.~Lemaire},
E.~Locci,
P.~Perez,
J.~Rander,
\mbox{J.-F.~Renardy},
A.~Roussarie,
\mbox{J.-P.~Schuller},
J.~Schwindling,
A.~Trabelsi,$^{21}$
B.~Vallage
\nopagebreak
\begin{center}
\parbox{15.5cm}{\sl\samepage
CEA, DAPNIA/Service de Physique des Particules,
CE-Saclay, F-91191 Gif-sur-Yvette Cedex, France$^{17}$}
\end{center}\end{sloppypar}
\vspace{2mm}
\begin{sloppypar}
\noindent
S.N.~Black,
J.H.~Dann,
R.P.~Johnson,
H.Y.~Kim,
N.~Konstantinidis,
A.M.~Litke,
M.A. McNeil,
G.~Taylor
\nopagebreak
\begin{center}
\parbox{15.5cm}{\sl\samepage
Institute for Particle Physics, University of California at
Santa Cruz, Santa Cruz, CA 95064, USA$^{22}$}
\end{center}\end{sloppypar}
\vspace{2mm}
\begin{sloppypar}
\noindent
C.N.~Booth,
S.~Cartwright,
F.~Combley,
M.S.~Kelly,
M.~Lehto,
L.F.~Thompson
\nopagebreak
\begin{center}
\parbox{15.5cm}{\sl\samepage
Department of Physics, University of Sheffield, Sheffield S3 7RH,
United Kingdom$^{10}$}
\end{center}\end{sloppypar}
\vspace{2mm}
\begin{sloppypar}
\noindent
K.~Affholderbach,
A.~B\"ohrer,
S.~Brandt,
C.~Grupen,
G.~Prange
\nopagebreak
\begin{center}
\parbox{15.5cm}{\sl\samepage
Fachbereich Physik, Universit\"at Siegen, D-57068 Siegen,
 Germany$^{16}$}
\end{center}\end{sloppypar}
\vspace{2mm}
\begin{sloppypar}
\noindent
G.~Giannini,
B.~Gobbo
\nopagebreak
\begin{center}
\parbox{15.5cm}{\sl\samepage
Dipartimento di Fisica, Universit\`a di Trieste e INFN Sezione di Trieste,
I-34127 Trieste, Italy}
\end{center}\end{sloppypar}
\vspace{2mm}
\begin{sloppypar}
\noindent
J.~Rothberg,
S.~Wasserbaech
\nopagebreak
\begin{center}
\parbox{15.5cm}{\sl\samepage
Experimental Elementary Particle Physics, University of Washington, WA 98195
Seattle, U.S.A.}
\end{center}\end{sloppypar}
\vspace{2mm}
\begin{sloppypar}
\noindent
S.R.~Armstrong,
E.~Charles,
P.~Elmer,
D.P.S.~Ferguson,
Y.~Gao,
S.~Gonz\'{a}lez,
T.C.~Greening,
O.J.~Hayes,
H.~Hu,
S.~Jin,
P.A.~McNamara III,
J.M.~Nachtman,$^{2}$
J.~Nielsen,
W.~Orejudos,
Y.B.~Pan,
Y.~Saadi,
I.J.~Scott,
J.~Walsh,
Sau~Lan~Wu,
X.~Wu,
G.~Zobernig
\nopagebreak
\begin{center}
\parbox{15.5cm}{\sl\samepage
Department of Physics, University of Wisconsin, Madison, WI 53706,
USA$^{11}$}
\end{center}\end{sloppypar}
}
\footnotetext[1]{Also at CERN, 1211 Geneva 23, Switzerland.}
\footnotetext[2]{Now at University of California at Los Angeles (UCLA),
Los Angeles, CA 90024, U.S.A.}
\footnotetext[3]{Also at Dipartimento di Fisica, INFN, Sezione di Catania, 
95129 Catania, Italy.}
\footnotetext[4]{Also Istituto di Fisica Generale, Universit\`{a} di
Torino, 10125 Torino, Italy.}
\footnotetext[5]{Also Istituto di Cosmo-Geofisica del C.N.R., Torino,
Italy.}
\footnotetext[6]{Supported by the Commission of the European Communities,
contract ERBFMBICT982894.}
\footnotetext[7]{Supported by CICYT, Spain.}
\footnotetext[8]{Supported by the National Science Foundation of China.}
\footnotetext[9]{Supported by the Danish Natural Science Research Council.}
\footnotetext[10]{Supported by the UK Particle Physics and Astronomy Research
Council.}
\footnotetext[11]{Supported by the US Department of Energy, grant
DE-FG0295-ER40896.}
\footnotetext[12]{Permanent address: Kangnung National University, Kangnung, 
Korea.}
\footnotetext[13]{Supported by the US Department of Energy, contract
DE-FG05-92ER40742.}
\footnotetext[14]{Supported by the US Department of Energy, contract
DE-FC05-85ER250000.}
\footnotetext[15]{Permanent address: Universitat de Barcelona, 08208 Barcelona,
Spain.}
\footnotetext[16]{Supported by the Bundesministerium f\"ur Bildung,
Wissenschaft, Forschung und Technologie, Germany.}
\footnotetext[17]{Supported by the Direction des Sciences de la
Mati\`ere, C.E.A.}
\footnotetext[18]{Supported by Fonds zur F\"orderung der wissenschaftlichen
Forschung, Austria.}
\footnotetext[19]{Now at SAP AG, 69185 Walldorf, Germany.}
\footnotetext[20]{Now at Harvard University, Cambridge, MA 02138, U.S.A.}
\footnotetext[21]{Now at D\'epartement de Physique, Facult\'e des Sciences de Tunis, 1060 Le Belv\'ed\`ere, Tunisia.}
\footnotetext[22]{Supported by the US Department of Energy,
grant DE-FG03-92ER40689.}
\footnotetext[23]{Now at Universit\'e de Lausanne, 1015 Lausanne, Switzerland.}
\footnotetext[24]{Now at INFN Sezione de Ferrara, 44100 Ferrara, Italy.}
%
%
\setlength{\parskip}{\saveparskip}
\setlength{\textheight}{\savetextheight}
\setlength{\topmargin}{\savetopmargin}
\setlength{\textwidth}{\savetextwidth}
\setlength{\oddsidemargin}{\saveoddsidemargin}
\setlength{\topsep}{\savetopsep}
\normalsize
\newpage
\pagestyle{plain}
\setcounter{page}{1}

\normalsize

\pagenumbering{arabic}
\pagestyle{plain}

\setcounter{page}{1}

%
%

\section{Introduction}

The study of $\tau$ decays involving kaons is necessary in order to
understand the strange hadronic $\tau$ sector as such, but also to
provide a better control over nonstrange decay modes. Both issues
play an important role in the low energy QCD analyses accessible through
hadronic $\tau$ decays. 

Results on charged kaons in three-prong decays~\cite{3prong} and
on $K^0_S$ production~\cite{k0decay} have already been published. In this
paper, one-prong decays are investigated. Taking advantage of their simpler
topology, the analysis combines the dE/dx measurement of the charged track 
and the reconstruction of $K^0_L$'s through their hadronic showers.   

The first analysis of one-prong final states with 
a charged kaon was presented using 38k selected $\tau^+\tau^-$ events 
collected in 1991 and 1992~\cite{1prong}, and an update was given
adding the 1993 data for a total of 64k selected $\tau^+\tau^-$
events~\cite{hadbrs}, enabling the study of 
the inclusive kaon mode and the three exclusive channels\footnote{Throughout 
this paper, charge conjugate states are implied.} $K^-\nu_\tau$,
$K^-\pi^0\nu_\tau$, and $K^-\pi^0\pi^0\nu_\tau$.
Analyses with $K^0_L$ detection were published with the same data 
sets~\cite{hadbrs,k0long}. In this paper, 
final results are given using the full LEP1 statistics corresponding to 
161k selected $\tau^+\tau^-$ events. 
Several improvements are implemented, 
increasing selection efficiencies and reducing systematic uncertainties.
This is achieved through a better use of all the available calorimetric
information, imbedded into a more efficient method for measuring the 
$K^0_L$ signal. Because of the larger statistics, a better understanding of 
the systematic effects is also acquired. The present results therefore
supersede those from previous analyses on one-prong kaon 
modes~\cite{1prong,hadbrs,k0long}.  

After a short description of the ALEPH detector and the event selection,
the dE/dx calibration using muons is presented in detail.
Then, the calorimetric method
for $K^0_L$ detection, taking advantage of the granularity of the 
calorimeters and the special $\tau$ decay kinematics, is described.
Branching ratios for inclusive charged kaon production and exclusive
channels involving charged and neutral kaons are derived through statistical
methods, both for dE/dx and calorimetry. A detailed account of the 
systematic uncertainties is given next, showing that in all cases the 
measurements are still statistically limited. Finally, the underlying 
dynamics in these decays is addressed through the study of invariant mass
spectra for exclusive decay samples.

\section{The ALEPH detector}

A detailed description of the ALEPH detector can be found 
elsewhere~\cite{aleph}. Here only the most relevant features for this analysis
are described.

The momenta of the charged tracks are measured by means of a central tracking 
system consisting of a silicon microstrip vertex detector, an inner tracking 
chamber (ITC), and a time projection chamber (TPC). All of these three
tracking devices are immersed in a 1.5 T axial magnetic field, and achieve a 
transverse momentum resolution for the charged track of   
$\sigma_{p_t}/p_t = 6 \times 10^{-4} p_t\bigoplus0.005$, with $p_t$ in GeV/c. 

Two subdetectors, the electromagnetic calorimeter (ECAL) and
the hadron calorimeter (HCAL), produce the energy measurement for
electrons and hadrons.
The ECAL consists of 45 layers of sandwiched lead sheets and 
proportional wire chambers, representing 22 radiation lengths, and 
is used to measure the energy deposit of electrons and photons. 
Its fine granularity is essential to the reconstruction of $\pi^0$'s, and 
the longitudinal segmentation with three stacks measures
the electromagnetic profile of the detected shower.
The energy resolution in the ECAL is measured to be 
$\sigma_E /E=0.18/ \sqrt{E\mbox{(GeV)}}+0.09$. 
Hadronic energy is measured by the
HCAL, which is composed of a 1.2 m thick iron stack 
serving as the yoke for magnetic flux return. Streamer tubes
are inserted into the gaps of the iron stacks, giving 23 energy samples.  
The energy resolution of the HCAL for pions
at normal incidence is $\sigma_E /E=0.85/ \sqrt{E\mbox{(GeV)}}$. 
Outside the HCAL, two layers of streamer tubes are used to complete
the muon detection.

In the data sample used for this analysis, all components of the
detector are required to be operational. Furthermore, one of the
following trigger conditions is required: a minimum ECAL energy of
6 GeV, or a track segment in the ITC pointing to an energy deposit in the
ECAL of at least 1.2 GeV, or a track segment matching the signal of a 
penetrating particle in the HCAL. By comparing redundant and independent
triggers involving the tracking detectors and the main calorimeters, 
the trigger efficiency for $\tau$ pair events is estimated to be better
than 99.99$\%$ within the selection criteria used in this analysis.

The measurement of the ionization dE/dx deposited by the
charged particles when traversing the TPC volume 
is the only tool in ALEPH for the identification of charged hadrons. 
Ideally, the dE/dx ionization can be measured with up to 338 samples.
However, in some case a charged track may not have such 
a number of samples because of the following factors: 
the cracks between the TPC sectors, the geometrically available track 
length and the applied truncation for avoiding the large fluctuations
caused by $\delta$ rays. A minimum of 40 samples is required
to ensure a good particle identification, 
rejecting about $0.1\%$ of the events. 

\section{Event selection}

A sample of about 161k events is obtained using a general 
$\tau$-pair selection procedure \cite{hadbrs,lbrs}, from
about 202k produced $\tau$ pairs.
The one-prong decays are selected from this sample.

Each event is first divided into two hemispheres separated by the plane 
perpendicular to the thrust axis of the event. A one-prong hemisphere
is required to contain exactly one good charged track, 
satisfying a minimum of four associated hits in the TPC, 
an impact parameter in the plane transverse to the beam axis 
$|d_0|\leq2$ cm and the distance from the interaction point 
along the beam axis $|z_0|\leq10$ cm. Also, the track momentum 
is required to be larger than 100 MeV/$c$. To reduce the number of 
fake $K^{0}_{L}$'s produced by additional tracks, each hemisphere 
candidate is not allowed to contain any unselected charged track 
associated with at least three hits either in the TPC or the ITC, 
$|d_0|\leq 20$ cm, $|z_{0}|\leq40$ cm and a momentum more than 2 GeV/$c$.   
Finally, to identify electrons, muons and 
hadrons in the one-prong $\tau$ candidates,
a particle identification~\cite{lbrs} is performed, except for 
the tracks with momenta below 2 GeV/$c$ 
(where only electrons are identified through their dE/dx) or
going through a crack between ECAL modules. All these requirements 
select a total sample of 143\,319 one-prong hadronic $\tau$ decays.
Hemispheres containing a track with momentum less than 
2 GeV/$c$ and not identified as an electron by dE/dx 
are kept for the $K^0_L$ search performed in Section~7.

Then, all selected one-prong candidates are classified according to 
the number of associated $\pi^0$'s, which are identified using a 
photon and $\pi^0$ reconstruction algorithm~\cite{hadbrs}. In this analysis, a
$\pi^0$ is defined as follows. A resolved $\pi^0$ 
is formed by a pair of photons with an invariant mass consistent with 
that of a $\pi^0$. Photons from high energy $\pi^0$'s may give rise to 
electromagnetic showers so close to each other that they cannot be recognized 
as two separate objects, leading to an unresolved $\pi^0$ which is identified 
by the energy-weighted moment technique~\cite{hadbrs}.
The remaining photons inside a cone of $30^\circ$ around the thrust axis
are called residual photons. Under these definitions,
one-prong $\tau$ decays are separated into 
$h^-$ ($h=\pi$ or $K$), $h^-\pi^0$, $h^-\pi^0\pi^0$, $h^-\pi^0\pi^0\pi^0$ 
and the rest~(containing one or several residual photons), 
corresponding to 47\,543, 54\,171, 10\,325, 959 and 30\,321 
decays, respectively.

Finally, to evaluate the efficiencies and the background contamination, 
1.0 million $\tau^+ \tau^-$ Monte Carlo events and 5.8 million $q\bar{q}$ 
are produced by the standard KORALZ~\cite{koralz} and the 
JETSET 7.4~\cite{jetset} generators, respectively, with a complete 
ALEPH detector simulation. 

\section{dE/dx calibration and probability functions}

The ionization energy deposit $R$ is measured in the TPC. The expected 
$R_i$ and its standard deviation $\sigma_i$ 
for a given particle type hypothesis $i$ are initially calculated using 
a parametrization of the Bethe-Bloch formula fitted to hadronic
events~\cite{aleph}, in which the environment is rather different from
that of the one-prong $\tau$ decay where no track overlap occurs.
In one-prong hadronic $\tau$ decays, the environment is similar to that of 
the muons from $\mbox{Z}\to\mu^+\mu^-$ or 
$\tau^-\to\mu^-\bar{\nu}_\mu\nu_\tau$ which provide
a pure sample of isolated muon tracks covering the full momentum range and
characterized by the same angular distribution in the detector as the hadron
sample under study. Therefore,
a special and accurate dE/dx calibration can be performed in order to
achieve a more precise measurement on the small kaon fraction in $\tau$ decays.

The main philosophy of dE/dx calibration follows 
Ref.~\cite{1prong}. Some modifications are made, taking into account 
slight variations in each period of data taking and allowing for
possible small differences between muons and hadrons, due to interactions
for the latter. 

\begin{figure}
\begin{center}
\vspace{-1.0cm}
\epsfxsize 10cm
\epsffile{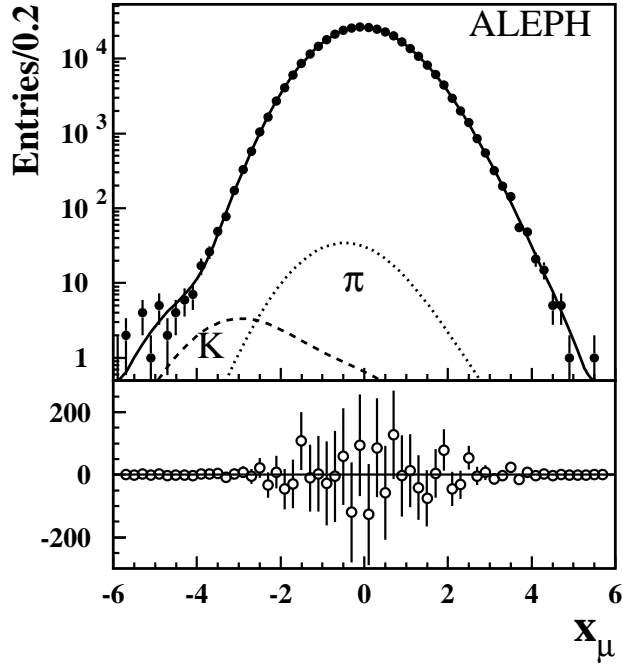}
\end{center}
\vspace{-3.0cm}
\caption{\it $x_\mu$ distribution for the muon sample (top).
The fit is indicated by the solid curve, and the expected residual 
kaons and pions are shown by dashed and 
dotted curves, respectively. The residuals between the data and
the fit are shown in the bottom plot.}
\label{muons}
\end{figure}

The calibration is based on a sample of muons selected from 
$\mbox{Z}\to \mu^+\mu^-$ and 
$\tau^{-}\to \mu^{-}\bar{\nu}_{\mu}\nu_{\tau}$, corresponding to
278\,079 and 49\,620 tracks respectively. Distributions
of the variable $x_{\mu}=\frac{R-R_{\mu}}{\sigma_{\mu}}$ are studied
as functions of (i) the period of data acquisition;
(ii) the polar angle of the track; (iii) the number of dE/dx samples; and 
(iv) the $R$ value. The calibration gives the new values for
$R_{\mu}$ and $\sigma_{\mu}$ according to
\begin{equation}
R^{new}_{\mu} = R^{old}_{\mu} + \bar{x}^{old}_{\mu}\times\sigma^{old}_{\mu}
\end{equation}
and
\begin{equation}
\sigma^{new}_{\mu} = \sigma_{x^{old}_\mu}\times\sigma^{old}_{\mu},
\end{equation}
where $\bar{x}^{old}_{\mu}$ and $\sigma_{x^{old}_\mu}$ represent
the mean and the standard deviation of the initial 
$x_{\mu}$ distribution. The calibration is performed in an iterative way
correcting most of the correlations among the above four factors.
The distribution of the calibrated $x_\mu$
for the muon sample is parametrized by an initially energy-independent 
four-Gaussian function 
$F_{\mu}(x)$ showing a slight asymmetry. 
There is still a tiny contamination from residual kaons 
and pions in the selected muon sample from $\tau$'s because of either hadron
misidentification as muon or decay in flight. These backgrounds are estimated
directly from the Monte Carlo simulation, and the corresponding 
fractions are fixed in the fit. The final result of the parametrization 
to the muon sample is shown in Fig.~\ref{muons}, indicating an excellent 
fit with $\chi^2/ndf=42.9/46$. 

The energy independence of $F_{\mu}(x)$ is not well satisfied 
at low momentum, as demonstrated with a large statistics sample of 
$\gamma\gamma\to\mu^+\mu^-$ events. It is observed that, in the region below 
7 GeV/$c$, the function becomes more symmetric. This small effect is taken 
into account in the final parametrization of the resolution function.
\begin{figure}
\begin{center}
\vspace{-2.0cm}
\epsfxsize 10cm
\epsffile{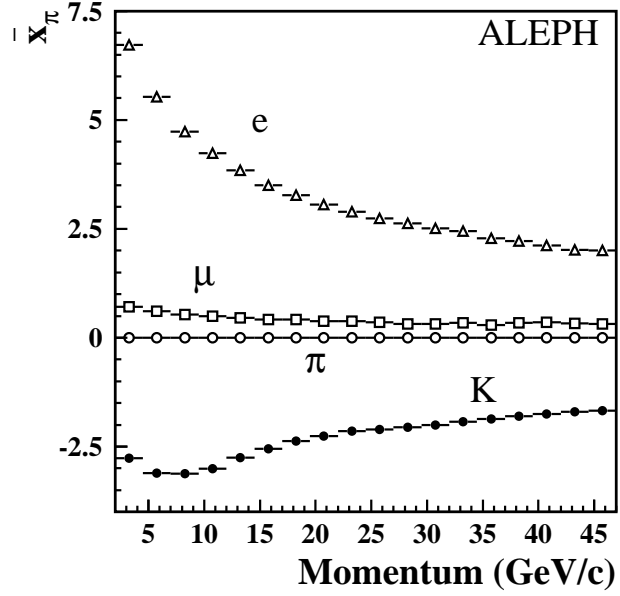}
\end{center}
\vspace{-0.5cm}
\caption{\it Separation from pions in standard deviations 
for different particle types as a function of momentum.}
\label{xpisep}
\end{figure}

Finally, the $x_h(h)$ distribution 
is treated to be of the same functional form as used for muons, namely
\begin{equation}
F_{h}(x_{h}) = F_{\mu}(x_{\mu}).
\end{equation}
With this function, the expected dE/dx response can be generated for
any particle type, resulting in a distribution from kaons in terms of $x_{\pi}$
of the form 
\begin{equation}
G_K(x_{\pi})=\frac{\sigma_{\pi}}{\sigma_{K}}
F_K\left(\frac{x_{\pi}\sigma_{\pi}+R_{\pi}-R_K}{\sigma_K}\right).
\label{fpi}
\end{equation}
The final particle separations in standard deviations are illustrated 
as a function of momentum in Fig.~\ref{xpisep}. 

\section{Calorimetric calibration}

Since the $K^0_L$ component is extracted from the calorimetric response by 
means of a statistical technique, a calibration of the measurements in both 
the ECAL and the HCAL is necessary. The details of the calorimetric 
calibration are described below.

\subsection{Definition of the discriminating variables 
$\delta\phi$ and $\delta_{E}$}

At LEP energies, the produced $\tau$ leptons are strongly Lorentz-boosted with
only a few degrees opening angle between any two particles in the 
final state. However, the bending due to the magnetic field provides an 
opportunity for separating the calorimetric clusters associated to the $K^0_L$
and the charged hadron in the HCAL, by making use of the transverse 
angular offset  
\begin{equation}
\delta\phi=\pm|\phi_{barycentre}-\phi_{track~impact}|,
\end{equation}
\noindent
where the positive sign is taken if the azimuthal angle 
$\phi_{barycentre}$, determined
by the barycentre of energy measured in the HCAL, is inside the bending 
trajectory of the charged hadron, and the negative sign is taken otherwise;
$\phi_{track~impact}$ is defined as the azimuthal angle for the position 
of the charged track at the inner radius of the HCAL. The minimum transverse 
momentum needed for a charged particle
to reach the HCAL is about 0.7 GeV/$c$ provided that no 
energy is lost in the ECAL. $K^0_L$'s from $\tau$ decay are emitted at small
angle with respect to the charged hadrons, leading to
a characteristic negative value of $\delta\phi$ in most of the cases.
Throughout this paper $\delta\phi$ is expressed in degrees.

In addition, the $K^0_L$ signal can be enhanced by exploiting the 
hadronic energy excess in the calorimeters, measured by the quantity
\begin{equation}
\delta_E = \frac{E - (E_{\pi^0})-P_h}{C\sqrt{P_h}},
\label{deltae}
\end{equation}
where $E$ is the sum of the energies measured by both
the ECAL and the HCAL in the studied hemisphere, $E_{\pi^0}$ is 
the total $\pi^0$ energy measured by the ECAL in $h\pi^0(\pi^0)$ modes,
$P_h$ is the charged track momentum, the denominator is close 
to the actual resolution of the calorimeters for
hadrons, and $C=0.9$ (GeV/$c$)$^{\frac{1}{2}}$. 
Since $K^0_L$'s from $\tau$ decay have a momentum 
in excess of 3.5 GeV/$c$, a positive value of $\delta_E$ is expected. 
This procedure differs from the previous 
analyses~\cite{1prong,hadbrs}, which used the energy deposit in the HCAL only.
In fact more than $50\%$ of $K^0_L$'s interact inside the 
ECAL and leave part of their energy mostly in the last 
two stacks, with longitudinal and transverse profiles different 
from those of a photon shower. Since the simulation of the relative energy
deposits in the ECAL and the HCAL is rather delicate, it is more 
reasonable to combine the two pieces of information, in order to reduce
systematic uncertainties.

In the following, cuts on $\delta\phi$ and $\delta_E$
are used separately in order to select samples for calibration. 

\subsection{Calibration for the energy measurement}

Decays with $\delta\phi>0$, consisting of single charged hadrons with a very 
small contamination from $K^0_L$ and unreconstructed 
$\pi^0$, are used for the energy calibration. The total hadronic energy 
measured by the calorimeters is summed with a 1.3 
weight factor for the ECAL, taking into account
the ratio of the electromagnetic calorimeter responses to electrons 
and pions~\cite{aleph}. The calibration provides 
corrections as a function of the data-taking period and 
the polar angle, the latter corresponding to varying effective interaction
length and geometry of the calorimeters. 
The same procedure is performed on the Monte Carlo
sample so that its $\delta_E$ distribution matches data. 

\begin{figure}
\begin{center}
\vspace{-2.0cm}
\epsfxsize 16cm
\epsffile{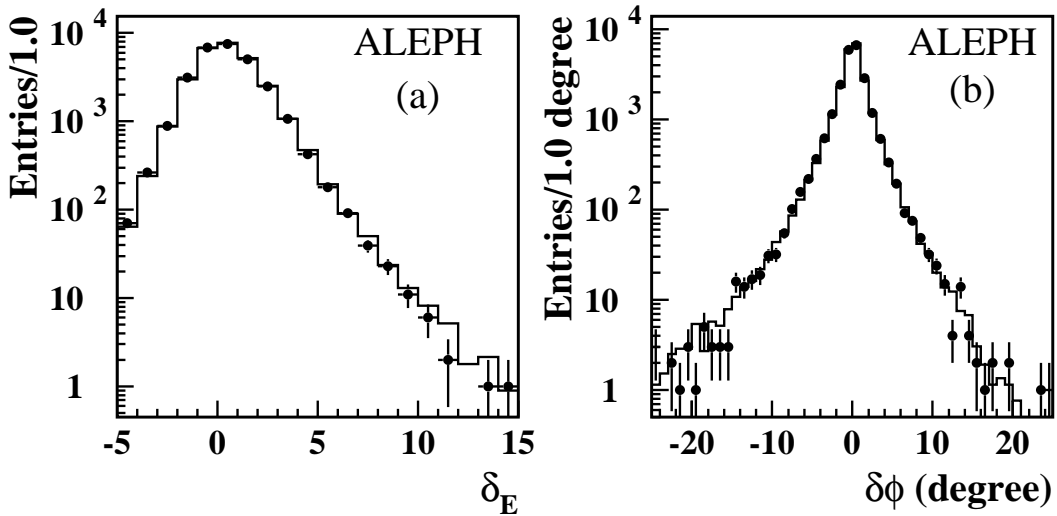}
\end{center}
\caption{\it (a) The calibrated $\delta_E$ distributions
for hemispheres with a positive $\delta\phi$. (b) 
The calibrated $\delta\phi$ distributions
for hemispheres with a negative  $\delta_E$. Data and Monte Carlo
simulation are shown in dots with error bars and in histograms,
respectively.}  
\label{calib}
\end{figure}

The calibration procedures are applied separately for 
the $h$, $h\pi^0$ and $h\pi^0\pi^0$ samples. 
Figure~\ref{calib}(a) shows a good agreement between data and Monte Carlo
simulation in the calibrated $\delta_E$ distributions for all the modes.
The above calibration is applied to all events, assuming that
events with a negative $\delta\phi$ follow the same variation as
those with a positive $\delta\phi$.

\subsection{Calibration for the barycentre offset $\delta\phi$ in the HCAL}

At this stage, a non-$K^0_L$ sample is selected by requiring
hemispheres with a negative sign of $\delta_E$. The $\delta\phi$ distribution
is checked and found to be consistent within different periods of 
data acquisition. For all hadronic samples, 
the Monte Carlo simulation is in good agreement
with data for the $\delta\phi$ mean, but produces a somewhat larger width.
This effect is proportional to the energy deposited in the HCAL, and is
used as a correction to the width of the $\delta\phi$ distribution 
as a function of energy in the Monte Carlo simulation.
Figure~\ref{calib}(b) shows that for the calibrated $\delta\phi$ 
distribution in the $h$ mode, Monte Carlo simulation is observed to 
agree with data. The same calibration is applied to      
the events with a positive $\delta_E$. 

\section{Channels involving charged kaons}

In this section, the inclusive one-prong $\tau$ decay
is first investigated to finalize the dE/dx calibration, 
and a measurement of the corresponding kaon fractions is presented. The dE/dx 
measurement is also exploited to study the exclusive one-prong $\tau$ decays, 
which are classified into $K^0_L$-reduced and $K^0_L$-enriched 
samples. Branching ratios are measured for 
all the exclusive modes $K^-$, $K^-\pi^0$, $K^-\pi^0\pi^0$, 
$K^-\pi^0\pi^0\pi^0$, $K^-K^0$, $K^-K^0\pi^0$ and $K^-K^0\pi^0\pi^0$.

\subsection{Charged kaon fraction in the inclusive mode}

\begin{figure}
\vspace{-1.5cm}
\begin{center}
\epsfxsize 8cm
\epsffile{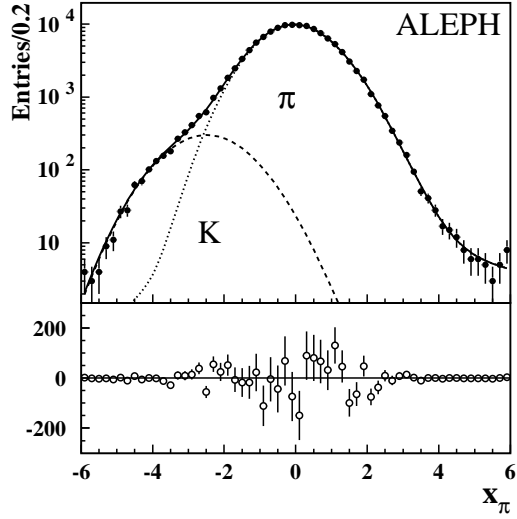}
\end{center}
\vspace{-1.5cm}
\caption{\it Fitted $x_{\pi}$ distribution for the inclusive 
one-prong $\tau$ decay sample (top). 
The dots with error bars correspond to data.
The fit, the expected pion (including the small background from electrons and
muons) and kaon contributions are indicated by the solid,
dotted and dashed curves, respectively. The goodness-of-fit for the 
overall $x_{\pi}$ distribution is $\chi^2/ndf=63/57$.
The residuals between the data and the fit are also shown in the bottom
plot.}
\label{inclk3}
\end{figure}

\begin{figure}
\vspace{-2.0cm}
\begin{center}
\epsfxsize 10cm
\epsffile{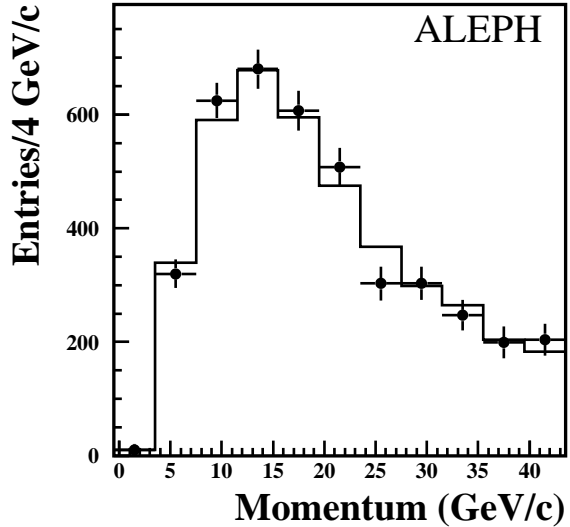}
\end{center}
\vspace{-1.0cm}
\caption{\it
The distribution of charged kaon momenta in the inclusive one-prong
$\tau$ decay. The dots with error bars correspond to the fitted data and
the histogram is for the Monte Carlo simulation.
Errors are statistical only.} 
\label{inclk4}
\end{figure}

To be insensitive to the $a~priori$ unknown dynamics, the fit to the $x_{\pi}$ 
distribution is performed in momentum slices. 
Small contaminations from residual muons and electrons are included
in the fit, with fractions estimated by the simulation.
The $K/\pi$ separation, varying with momentum as shown 
in Fig.~\ref{xpisep}, is determined by Monte Carlo and kept fixed in the fit. 
The $x_{\pi}(\pi)$ parameters and the kaon fraction are free to vary. 
Table~\ref{inclk1} lists the fit results in the
different momentum slices. Figures~\ref{inclk3} and~\ref{inclk4} 
show the sum of the fits to $x_{\pi}$ for all momentum 
slices and the kaon yield as a function of momentum for both 
data and simulation. 
 
Due to the large statistics, 
the inclusive $x_\pi$ fits yield the final calibration constants
(relative to the muon calibration) as a function of momentum 
which are used in the later fits to the $x_{\pi}$ distributions 
for all the exclusive modes. For the whole momentum range, the obtained values 
$\bar{x}_\pi=-0.021\pm0.004$ and $\sigma_{x_\pi}=0.992\pm0.003$ are 
significantly different from those (with similar errors) 
from the $\tau\to\mu$ sample which covers the same momentum range.
These shifts are consistent with those found using a 99.5$\%$ pure 
pion data sample in the $h^-\pi^{0}$ mode 
by requiring the invariant mass $M(K^-\pi^0)\geq 1.1$ GeV/$c^2$, in order to
veto the $K^*(892)^-$ background. The values
$\bar{x}_{\pi}=-0.02\pm0.01$ and $\bar{\sigma}_{\pi}=0.98\pm0.02$
are obtained in this case.

In estimating the efficiency for the inclusive mode, 
relative contributions from all exclusive decay modes are fixed in the Monte 
Carlo simulation with the exclusive branching ratios 
obtained later in this analysis, giving an efficiency 
value of $(64.05\pm0.03)\%$.
The branching ratio for the inclusive mode is therefore obtained
\begin{equation}
B(\tau^-\to K^- X\nu_{\tau}) = (1.52\pm0.04_{stat})\%,
\end{equation}
where $X$ represents any system of neutral particles (including 
$K^0_S$'s).

\begin{table}
\begin{center}
\small
\begin{tabular}{|l|c|c|c|c|c|c|}\hline\hline
$p$ (GeV/$c$)&$\bar{x}_{\pi}(\pi)$ & $\sigma_{\pi}(\pi)$  
& $f_K(\%)$ & $N_K$ & $N_{q\bar{q}}$ &$\chi^2$/ndf \\ \hline
$~2.0 - 3.5$&$-0.03\pm0.01$ &$1.00\pm0.01$ 
& $0.11\pm0.12(0.11)$ & $10\pm10$ & $6\pm2$ & 49.4/47\\ \hline
$~3.5 - 7.5$&$-0.03\pm0.01$ &$1.00\pm0.01$ 
& $1.30\pm0.11(0.10)$ & $320\pm25$ & $6\pm2$ &69.5/57\\ \hline
$~7.5 -11.5$&$-0.02\pm0.01$ &$0.99\pm0.01$ 
& $2.82\pm0.15(0.14)$ & $625\pm31$ & $4\pm2$ &73.6/53\\ \hline
$11.5 -15.5$&$-0.02\pm0.01$ &$0.99\pm0.01$
& $3.92\pm0.23(0.20)$ & $680\pm34$ & $8\pm2$ &49.3/50\\ \hline
$15.5 -19.5$&$-0.03\pm0.01$ &$0.98\pm0.01$ 
& $4.59\pm0.34(0.26)$ & $607\pm35$ & $1\pm1$ &38.6/49\\ \hline
$19.5 -23.5$&$-0.01\pm0.01$ &$0.98\pm0.01$ 
& $4.83\pm0.44(0.32)$ & $508\pm33$ & $7\pm2$ &38.6/42\\ \hline
$23.5 -27.5$&$-0.03\pm0.01$&$1.01\pm0.01$ 
& $3.48\pm0.50(0.35)$ & $303\pm30$ & $5\pm2$ &63.7/45\\ \hline
$27.5 -31.5$&$-0.02\pm0.02$ &$0.99\pm0.01$ 
& $4.17\pm0.58(0.39)$ & $303\pm29$ & $5\pm2$ &46.3/39\\ \hline
$31.5 -35.5$&$-0.02\pm0.02$ &$0.98\pm0.02$ 
& $4.30\pm0.69(0.46)$ & $247\pm27$ & $7\pm2$ &42.5/38\\ \hline
$35.5 -39.5$&$-0.02\pm0.02$ &$0.98\pm0.02$ 
& $4.35\pm0.99(0.60)$ & $199\pm28$ & $5\pm2$ &38.5/36\\ \hline 
$\geq 39.5$&$-0.02\pm0.02$ &$0.98\pm0.02$ 
& $5.04\pm1.21(0.70)$ & $204\pm28$ & $16\pm3$ &53.4/39\\ \hline\hline
\end{tabular}
\caption{\it Charged kaon momentum range, 
the fitted $x_{\pi}(\pi)$ parameters
and kaon fractions, kaon yields, estimated non-$\tau$ background 
together with the test of goodness of fit.
The errors in parentheses are those without the contribution
from the uncertainties on the $x_{\pi}(\pi)$ parameters.}
\label{inclk1}
\end{center}
\end{table}

\subsection{Selection for the $K^0_L$-reduced and $K^0_L$-enriched samples}

The analysis of exclusive modes involving kaons divides 
the one-prong $\tau$ decays into $K^0_L$-reduced and 
$K^0_L$-enriched samples, by using cuts on the two variables $\delta\phi$ and 
$\delta_E$. In the $K^0_L$-enriched samples, the net energy deposited 
in the ECAL should be at most a factor of two larger than in the HCAL, 
to remove the background hemispheres with unreconstructed $\pi^0$'s. 
Meanwhile, the requirements of $\delta\phi< 0$, $\delta_E\geq1.0$ and
$\delta\phi^2+\delta_E^2\geq 12$ are placed on the events to reduce the
non-$K^0_L$ background. Also, 
$K^-K^0\pi^0(\pi^0)$ candidates 
should be consistent with $\tau$ decay, 
$i.e.$, $M(K^-\pi^0)$ in  $K^-K^0\pi^0\nu_{\tau}$ or 
$M(K^-\pi^0\pi^0)$ in $K^-K^0\pi^0\pi^0\nu_{\tau}$ 
less than $(M_{\tau}-M_{K^0})$. 
To further suppress the charged pion background from the $\pi^-\pi^0(\pi^0)$ 
modes, the angle $\alpha_{open}$ between the direction 
pointing from the interaction point to the energy-weighted 
barycentre in the HCAL and 
the $\pi^0(\pi^0)$ momentum vector, is required to be smaller than 
$6^\circ$ in selecting the $K^-K^0\pi^0(\pi^0)$ candidates. 
Finally, the $K^-K^0\pi^0\pi^0$ candidates are required to 
have $M(\pi^0\pi^0)$ greater than 0.5 GeV/$c^2$,
to reduce the $K^-K^0\pi^0$ contamination due to the decay $K^0\to\pi^0\pi^0$. 

The above selection criteria define the $K^0_L$-enriched samples for
$K^-K^0$, $K^-K^0\pi^0$ and $K^-K^0\pi^0\pi^0$ in
the $h^-$, $h^-\pi^0$ and $h^-\pi^0\pi^0$ modes. The $K^0_L$-reduced
samples, involving decay modes like $K^-$, $K^-\pi^0$ and $K^-\pi^0\pi^0$,
are defined as those which do not satisfy these 
criteria. Finally, the kaon fraction for each sample is determined
from the fit to the corresponding $x_{\pi}$ distribution.

\subsection{Charged kaon fractions in the $K^0_L$-reduced samples}

\begin{figure}
\begin{center}
\vspace{-2.0cm}
\epsfxsize 16cm
\epsffile{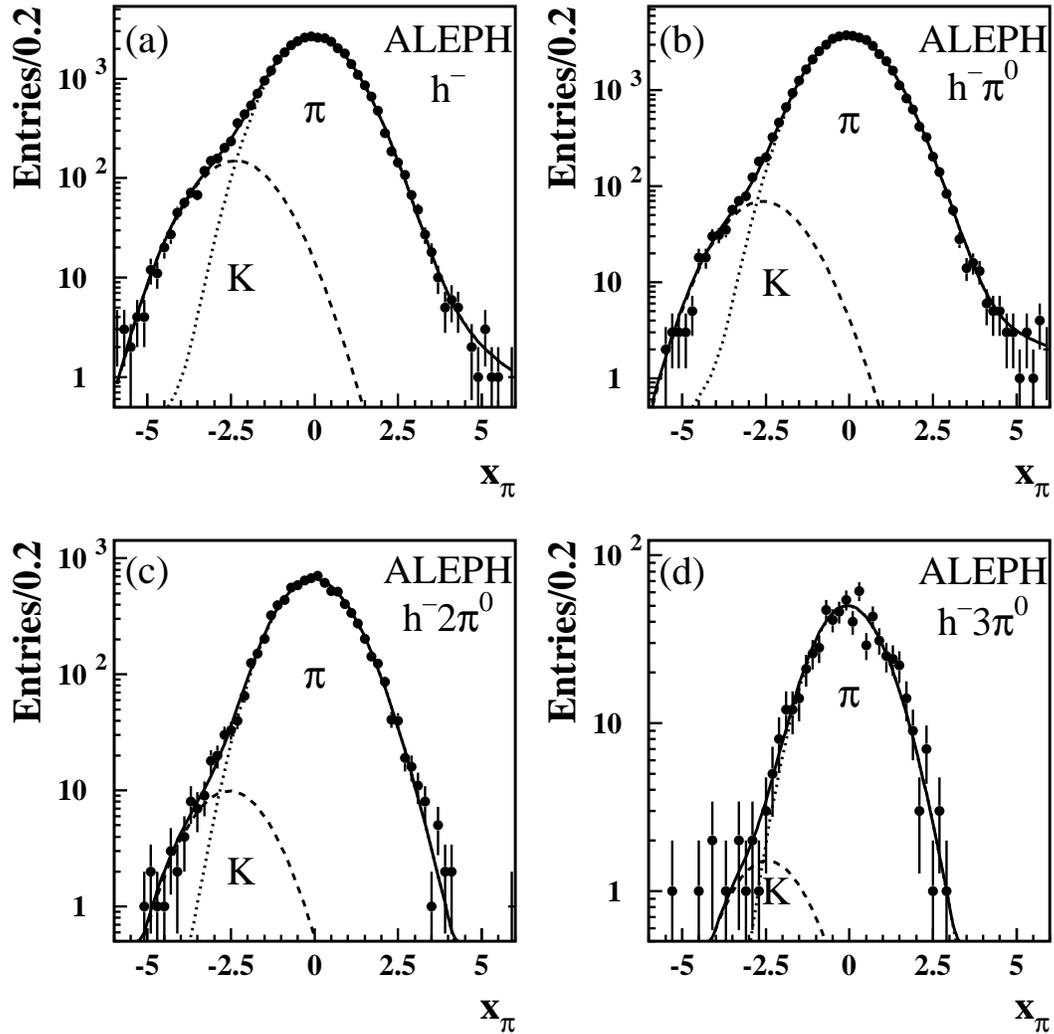}
\end{center}
\caption{\it The $x_{\pi}$ distributions in the $K^0_L$-reduced samples. Data
are shown with error bars. The fit results are shown as solid curves; 
the kaon and pion components are given by the dashed and dotted curves,
respectively; the small electron and muon contributions are included
in the pion component.}
\label{exclknp0}
\end{figure}

\begin{figure}
\begin{center}
\vspace{-2.0cm}
\epsfxsize 14cm
\epsffile{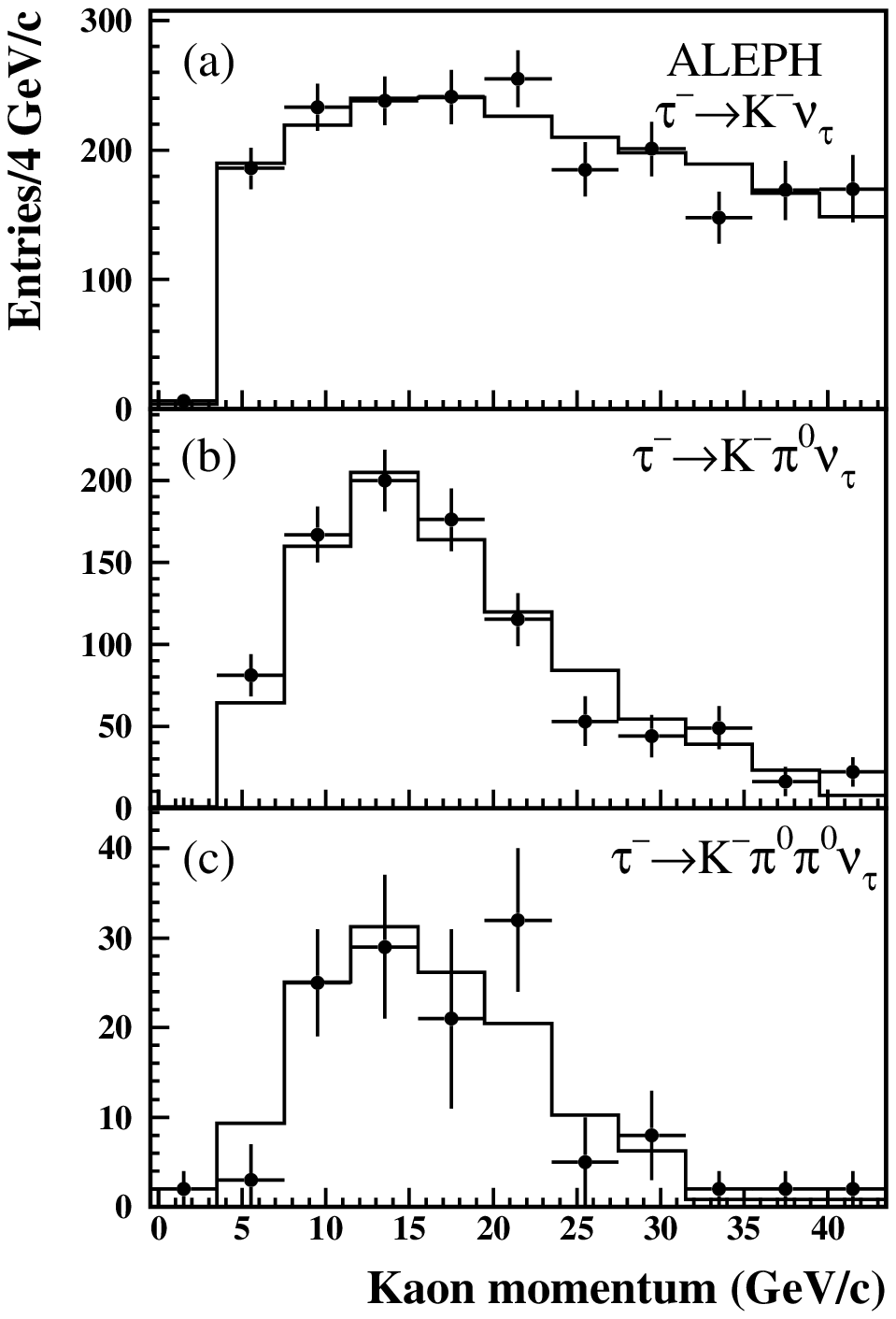}
\end{center}
\caption{\it Kaon momentum distributions for the $K^0_L$-reduced 
samples: (a) $h^-$ mode; (b) $h^-\pi^0$ mode; 
(c) $h^-\pi^0\pi^0$ mode. Data and Monte Carlo simulations are
shown in dots with errors and in histograms, respectively.
Errors on the measured kaon yields are statistical only.} 
\label{momk}
\end{figure}

The $K^0_L$-reduced samples $h^-$, $h^-\pi^0$ and $h^-\pi^0\pi^0$ are
used to measure the branching ratios for $K^-\nu_\tau$,
$K^-\pi^0\nu_\tau$ and $K^-\pi^0\pi^0\nu_\tau$.
Fits to the $x_\pi$ distributions (Fig.~\ref{exclknp0}) 
are performed in each momentum bin,
in order to reduce uncertainties from dynamics in the simulation.
The $x_{\pi}(\pi)$ parameters in each momentum bin are directly taken 
from the values obtained in the fits to the inclusive mode as given in 
Table~\ref{inclk1}. Tiny fractions of muon and electron backgrounds
are directly estimated from Monte Carlo and fixed in the fits.
Table~\ref{exclk} gives the 
results of the fits. Figure~\ref{momk} shows the charged kaon momentum
distributions for both data and Monte Carlo simulations using the 
branching ratios obtained later. Good agreement is observed. 
\begin{table}
\begin{center}
\begin{tabular}{|l|c|c|c|c|c|}\hline\hline
Mode & $f_K(\%)$ &$N_K$ &$\chi^2/ndf$ & $N_{q\bar{q}}$ 
&$B (10^{-3}$)\\ \hline
$K^-$ 
&$5.75\pm0.19$&$2032\pm66$&71.9/57&$11\pm4$&$6.96\pm0.25$\\ \hline
$K^-\pi^0$ 
&$1.92\pm0.10$&$ 923\pm47$&68.3/57&$15\pm5$&$4.44\pm0.26$\\ \hline
$K^-\pi^0\pi^0$ 
&$1.55\pm0.23$&$ 131\pm19$&52.0/47&$ 3\pm1$&$0.56\pm0.20$\\ \hline  
$K^-\pi^0\pi^0\pi^0$
&$3.43\pm1.09$&$  22\pm 7$&33.5/35&$<1 $&$0.37\pm0.21$\\ \hline \hline 
$K^-K^0$ 
&$9.20\pm0.96$&$ 150\pm14$&27.2/27&$ 2\pm1$&$1.62\pm0.21$\\ \hline  
$K^-K^0\pi^0$ 
&$9.23\pm1.18$&$  78\pm10$&19.7/20&$ 1\pm1$&$1.43\pm0.25$\\ \hline  
$K^-K^0\pi^0\pi^0$ 
&$8.89\pm4.44$&$   4\pm 2$&3.2/6&$ <1$&$<0.18~(95\%~\mbox{C.L.})$\\ 
\hline\hline  
\end{tabular}
\caption{\it Summary of statistics, overall 
charged kaon fractions in the $x_\pi$ fit,
the test of goodness-of-fit for the overall $x_\pi$ distribution in each mode,
estimated $q\bar{q}$ backgrounds, 
and the branching ratios obtained later. Errors are statistical only.}
\label{exclk}
\end{center}
\end{table}

For the $h^-\pi^0\pi^0\pi^0$ mode, no cut is applied on the calorimetric
measurement, and the kaon momentum is required to be greater than
5 GeV/$c$, reducing the pion background. The fit is only done for the whole
momentum region because of low statistics. A generator with a $V-A$ matrix 
element and $\tau$ phase space factor is 
used to produce Monte Carlo events for the decay 
$\tau^-\to K^-\pi^0\pi^0\pi^0\nu_{\tau}$, giving an estimate of
the average $K/\pi$ separation. A fit to the corresponding $x_{\pi}$ 
distribution is performed, and the result is given in Table~\ref{exclk}. 
Figure~\ref{exclknp0}(d) shows the 
$x_{\pi}$ distributions and the corresponding fits in the 
$K^-\pi^0\pi^0\pi^0$ mode.
  
\subsection{Charged kaon fractions in the $K^0_L$-enriched samples}

Decay modes such as $K^-K^0$, $K^-K^0\pi^0$
and $K^-K^0\pi^0\pi^0$ are measured by fitting the $x_{\pi}$ 
distributions in the $K^0_L$-enriched samples. 

\begin{figure}
\begin{center}
\vspace{-2.0cm}
\epsfxsize 16cm
\epsffile{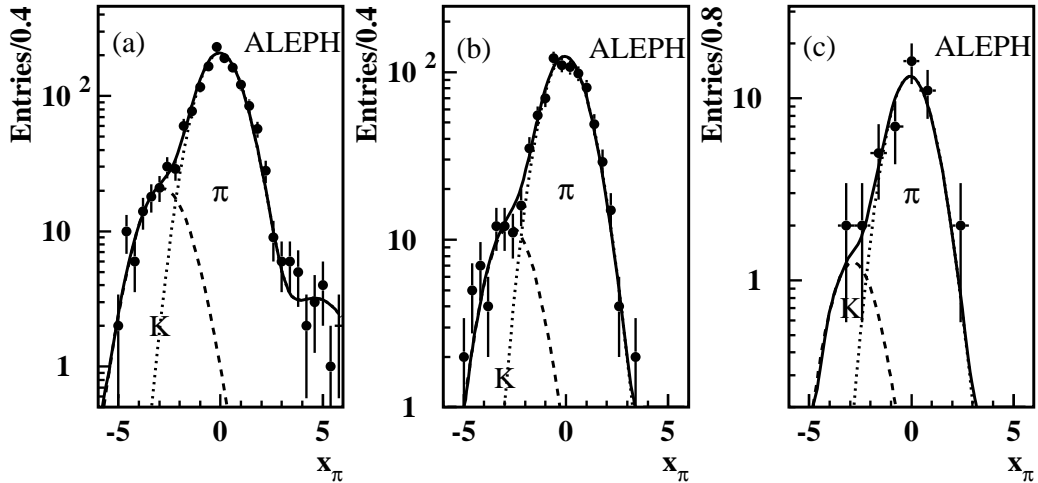}
\end{center}
\caption{\it Fitted $x_{\pi}$ distributions in the $K^0_L$-enriched
samples: (a) for $h^-K^0$, (b) for $h^-K^0\pi^0$ and (c) 
for $h^-K^02\pi^0$. The dots with error bars correspond to data.
The fit function and the pion (including the small electron and muon 
background) and kaon contributions are shown in solid, dotted and 
dashed, respectively.} 
\label{kk0np0}
\end{figure}

\begin{figure}
\begin{center}
\vspace{-2.0cm}
\epsfxsize 10cm
\epsffile{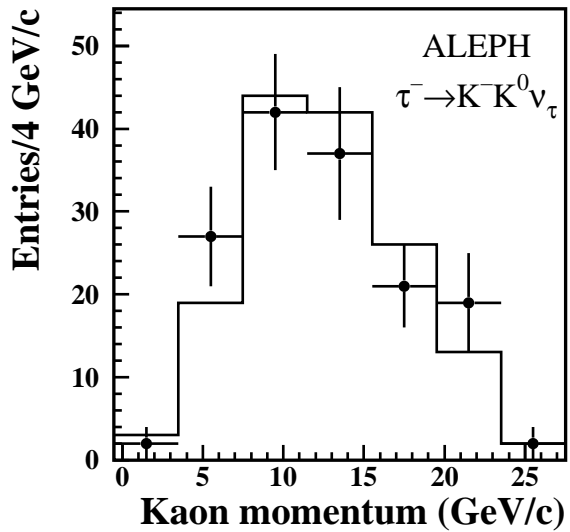}
\end{center}
\caption{\it Charged kaon momentum distributions for data (dots with
error bars) and the expected signal from simulation 
(histogram).}
\label{momkk0}
\end{figure}

In the analysis of $h^-K^0$ mode, the fits to the $x_{\pi}$ distributions 
in the $K^0_L$-enriched sample are also performed in momentum bins.
This sample contains the tracks going
through the ECAL cracks in order to enhance the selection efficiency,
consequently introducing some electron background which is
taken into account in the fits and has no influence on the kaon content. 
The fitting procedure is the same as used above, yielding the results
shown in Table~\ref{exclk} and Fig.~\ref{kk0np0}.
The charged kaon momentum distribution in the $h^-K^0$ mode is also shown in 
Fig.~\ref{momkk0}, indicating a good agreement between data and the Monte 
Carlo simulation.

Fits to the full momentum range are applied for 
the $h^-K^0\pi^0$ and $h^-K^0\pi^0\pi^0$ modes because of low statistics.
To reduce the pion background, an additional 5 GeV/$c$ cut is employed 
for the minimum kaon momentum. The average $K/\pi$ separation is obtained from 
the simulations. Table~\ref{exclk} and Fig.~\ref{kk0np0} 
show the results of the fits. 

\subsection{Branching ratios for decays with charged kaons}

The measurement of the kaon yields for all exclusive modes is used to 
determine the corresponding branching ratios after applying 
the efficiency matrix obtained from Monte Carlo (see Table~\ref{effk}).
In the simulation, form factors for the two- and three-meson decay channels 
are derived from Ref.~\cite{was,fink}, and the four-meson 
$\tau$ decay channels, for which no model is available at present, 
are generated according to the general $V-A$ weak
matrix element folded with phase space distributions for hadrons.
\begin{table}
\begin{center}
\footnotesize
\begin{tabular}{|c|c|c|c|c|c|c|c|c|}\hline\hline
Mode & $K^-$ & $K^-\pi^0$ & $K^-\pi^0\pi^0$ & $K^-\pi^0\pi^0\pi^0$ &
$K^-K^0$ & $K^-K^0\pi^0$ & $K^-K^0\pi^0\pi^0$ &$K^-\eta$\\ \hline
$K^-$ 
& 65.37 & 4.25 & 0.34 &$\sim0$  & 15.24 & 1.28 & 0.14 & 0.28  \\ \hline
$K^-\pi^0$
& 0.42 & 44.26 & 7.41 & 1.18 & 4.69 & 8.89 & 2.86 & 1.84 \\ \hline
$K^-\pi^0\pi^0$ 
&$\sim0$ & 0.50 & 23.32 & 12.40 & 2.90 & 3.69 & 6.30 & 5.34 \\ \hline
$K^-\pi^0\pi^0\pi^0$
&$\sim0$&$\sim0$  & 0.28 & 8.10 &$\sim0$ & 1.26 & 2.82& 1.82 \\ \hline\hline
$K^-K^0$
& 0.55 & 0.66 &$\sim0$ &$\sim0$ & 16.86 & 1.82 & 0.48&0.26 \\ \hline
$K^-K^0\pi^0$
&$\sim0$ & 0.48 & 0.63 &$\sim0$ & 1.32 & 10.04 & 3.74& 0.20 \\ \hline 
$K^-K^0\pi^0\pi^0$
&$\sim0$ &$\sim0$ &$\sim0$ & 0.42 &$\sim0$  & 0.62 & 4.54&0.24 \\ \hline\hline
\end{tabular}
\caption{\it Efficiency matrix (in percent) 
for the exclusive $\tau$ decay modes 
involving charged kaons. The $K^0_L$-reduced samples involve the first
four decay modes, while the $K^0_L$-enriched samples include the last three. 
The generated decay modes are given in the 
first row, and the reconstructed modes in the first column.}
\label{effk}
\end{center}
\end{table}

All channels studied in this analysis are correlated through
$\pi^0$ reconstruction and $K^0$ contamination.
Their branching ratios should thus be computed 
globally, by solving 
the seven linear equations with the efficiency matrix given in 
Table~\ref{effk}, in order to take the correlations properly 
into account. Since the channel $K^-\eta\nu_{\tau}$ can also
feed into each studied mode, the corresponding subtraction
is performed using the measurement 
$B(\tau^-\to K^-\eta\nu_{\tau})=(2.9^{\,+1.5}_{\,-1.4})\times 10^{-3}$ 
from ALEPH~\cite{keta1}. The final branching ratios for 
one-prong $\tau$ decays involving charged kaons are given in Table~\ref{exclk}.

\section{Channels involving $K^0_L$ only}

This section studies the $\overline{K^0}\pi^-$, 
$\overline{K^0}\pi^-\pi^0$ and $\overline{K^0}\pi^-\pi^0\pi^0$ decay modes, 
in which the potentially large backgrounds expected from the decay 
channels $\pi^-\nu_{\tau}$, $\pi^-\pi^0\nu_{\tau}$ and 
$\pi^-\pi^0\pi^0\nu_{\tau}$ are separated
by means of a fit method based on the previously defined variables
$\delta_E$ and $\delta\phi$.

\subsection{Selection for the $K^0_L$ samples}

In general, non-$K^0_L$ background can mimic a $K^0_L$ 
signal through (i) $h^-$ with upward
fluctuation of calorimetric energy; (ii) $\pi^0$ leakage into the HCAL and
(iii) unreconstructed $\pi^0$'s in the ECAL. 
As done in selecting the $K^0_L$-enriched samples for the purpose
of measuring the charged kaons, the selection criteria are based 
on the measurements from both the HCAL and the ECAL. The $K^0_L$ candidates
are required to have at least 5 GeV energy deposited in the HCAL, to
remove the background from a high energy $\pi^0$ leakage. The net 
energy deposited in the ECAL must be smaller than 15 GeV and less than two 
times the energy deposited in the HCAL, rejecting background with 
unreconstructed $\pi^0$'s. According to the Monte Carlo study, 
the $\pi^-\pi^0$ or $\pi^-\pi^0\pi^0$ backgrounds may produce
a fake $K^0_L$ signal in the HCAL when part of $\pi^0$ energy leaks into the 
HCAL through the cracks between ECAL modules. To remove this background
a quantity $\phi_{module}$, defined as the azimuthal angle for 
the energy-weighted barycentre in the HCAL,
is required to be outside a 5$^\circ$ window centred
on each ECAL crack. A further suppression of this
background is achieved by requiring that $\delta_E$ be 
larger than 1 for $\overline{K^0}\pi^-$ and 1.5 for 
$\overline{K^0}\pi^-\pi^0(\pi^0)$, at the cost of a few percent signal loss. 
The charged hadron momentum
is required to be less than 30 GeV/$c$ for $K^0\pi^-$, 25 GeV/$c$ for 
$\overline{K^0}\pi^-\pi^0$ and 20 GeV/$c$ for 
$\overline{K^0}\pi^-\pi^0\pi^0$. To be consistent with $\tau$ decay,
the $\overline{K^0}\pi^-\pi^0$ or $\overline{K^0}\pi^-\pi^0\pi^0$
candidates must satisfy $M(\pi^-\pi^0)$
or $M(\pi^-\pi^0\pi^0)$ less than $(M_\tau-M_{K^0})$.
The angle $\alpha_{open}$ 
defined in Section 6.2 is required to be smaller
than 8$^\circ$ for the $\overline{K^0}\pi^-\pi^0$ mode and 
5$^\circ$ for the $\overline{K^0}\pi^-\pi^0\pi^0$ mode.
The above cuts select three samples for the study of the 
$\overline{K^0}\pi^-$, $\overline{K^0}\pi^-\pi^0$ and 
$\overline{K^0}\pi^-\pi^0\pi^0$ modes.

\subsection{Measured $K^0_L$ fractions} 
 
\begin{figure}
\begin{center}
\vspace{-2.0cm}
\epsfxsize 14cm
\epsffile{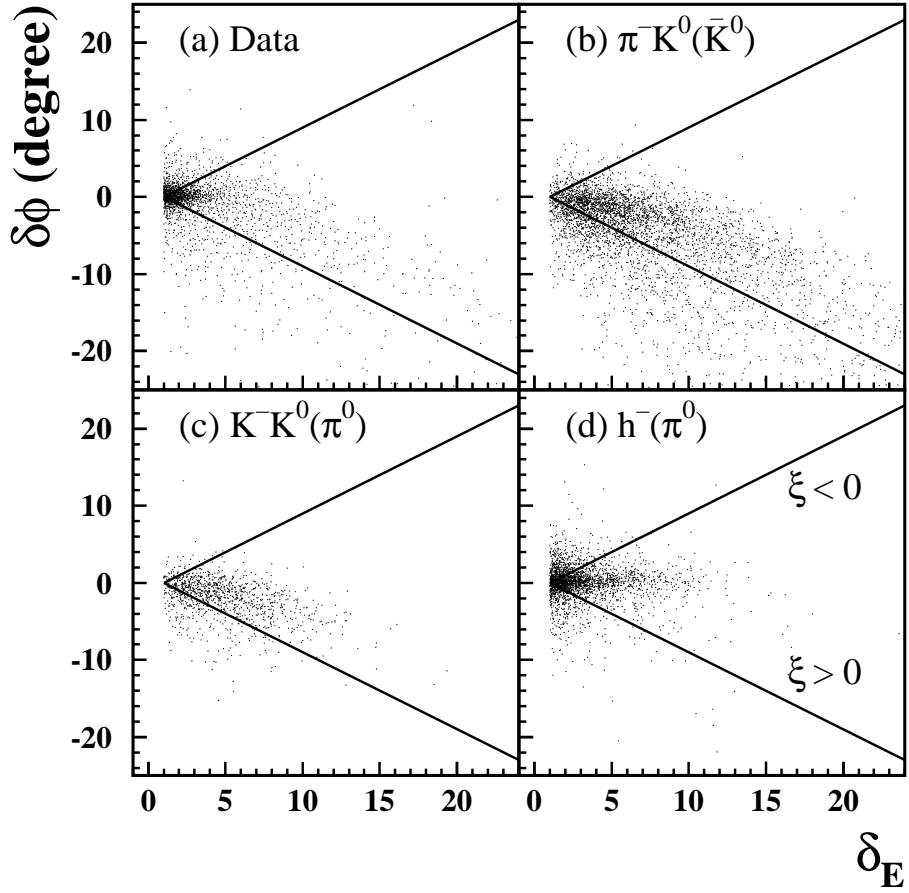}
\end{center}
\caption{\it The scatter plots ($\delta_E$ vs $\delta\phi$) in
$h$ mode for (a) data, (b) Monte Carlo events for 
$\tau^-\to \overline{K^0}\pi^-(K^0)\nu_{\tau}$ (${\sim}5\times$ 
statistics of data), (c) Monte Carlo events for 
$\tau^-\to K^-K^0(\pi^0)\nu_{\tau}$ (${\sim8}\times$
data), and (d) $\tau^-\to h^-(\pi^0)\nu_{\tau}$
(${\sim5}\times$ data). The two $\xi$ axes considered in the analysis 
are drawn with straight lines as shown in each plot.} 
\label{dedphi}
\end{figure}

The two variables $\delta_E$ and $\delta\phi$
are used to distinguish the $K^0_L$ events from the background. 
In previous analyses~\cite{1prong,hadbrs}, the $K^0_L$ signals are
selected with cuts on $\delta_E$ and $\delta\phi$, while 
in this analysis a fit method is applied. It is observed that
systematic effects from the dynamics are reduced, and at the same 
time the distributions of $\delta_E$ and $\delta\phi$ in data and
in the simulation can be compared, providing a way to estimate
the systematic uncertainties. Rather than performing 
a two-dimensional fit, a one-dimensional fit is exploited using the
variable $\xi$ 
\begin{equation}
\xi=-\frac{\sqrt{2}}{2}(\delta_E - \Delta+\delta\phi)
~~~~(\mbox{for}~\delta_E>0,~\delta\phi>0),
\end{equation}
and 
\begin{equation}
\xi=\frac{\sqrt{2}}{2}(\delta_E - \Delta-\delta\phi)
~~~~(\mbox{for}~\delta_E>0,~\delta\phi\leq0),
\end{equation}  
where $\Delta$ represents the $\delta_E$ cut used for selecting 
the $K^0_L$ sample. This procedure is equivalent to projecting each
event onto one of the two $\xi$ axes shown in 
Fig.~\ref{dedphi} together with scatter plots for data and Monte Carlo.
Large $\delta\phi$ and $\delta_E$ offsets
are observed for the $\overline{K^0}\pi^-(K^0)$ events, while they are 
relatively smaller in the $K^-K^0(\pi^0)$ sample, because of the harder
charged kaon momentum. The $h^-(\pi^0)$ backgrounds  
contribute to the $K^0_L$-enriched sample because of an upward 
energy fluctuation, yielding a long tail in the $\delta_E$ distribution. 
The definition of $\xi$ is optimized for the $\overline{K^0}\pi^-(K^0)$ 
signal. Events with a positive (negative) sign of 
$\delta\phi$ always correspond to a negative (positive) sign of $\xi$. As a
result, the majority of $K^0_L$'s are distributed in the region of large 
positive $\xi$ and are thus efficiently separated from the 
non-$K^0_L$ background.

To measure the $K^0_L$ fractions, the shapes of $\xi$ distributions for
both the $K^0_L$ modes and the non-$K^0_L$ modes are parametrized from 
the Monte Carlo simulation, using a set of Gaussian functions.
Since the majority of $K^0_L$ events are on the positive $\xi$ side,
the negative $\xi$ distribution can be used to check
the calorimetric calibration and the simulation as well. 

\begin{figure}
\begin{center}
\vspace{-2.0cm}
\epsfxsize 14cm
\epsffile{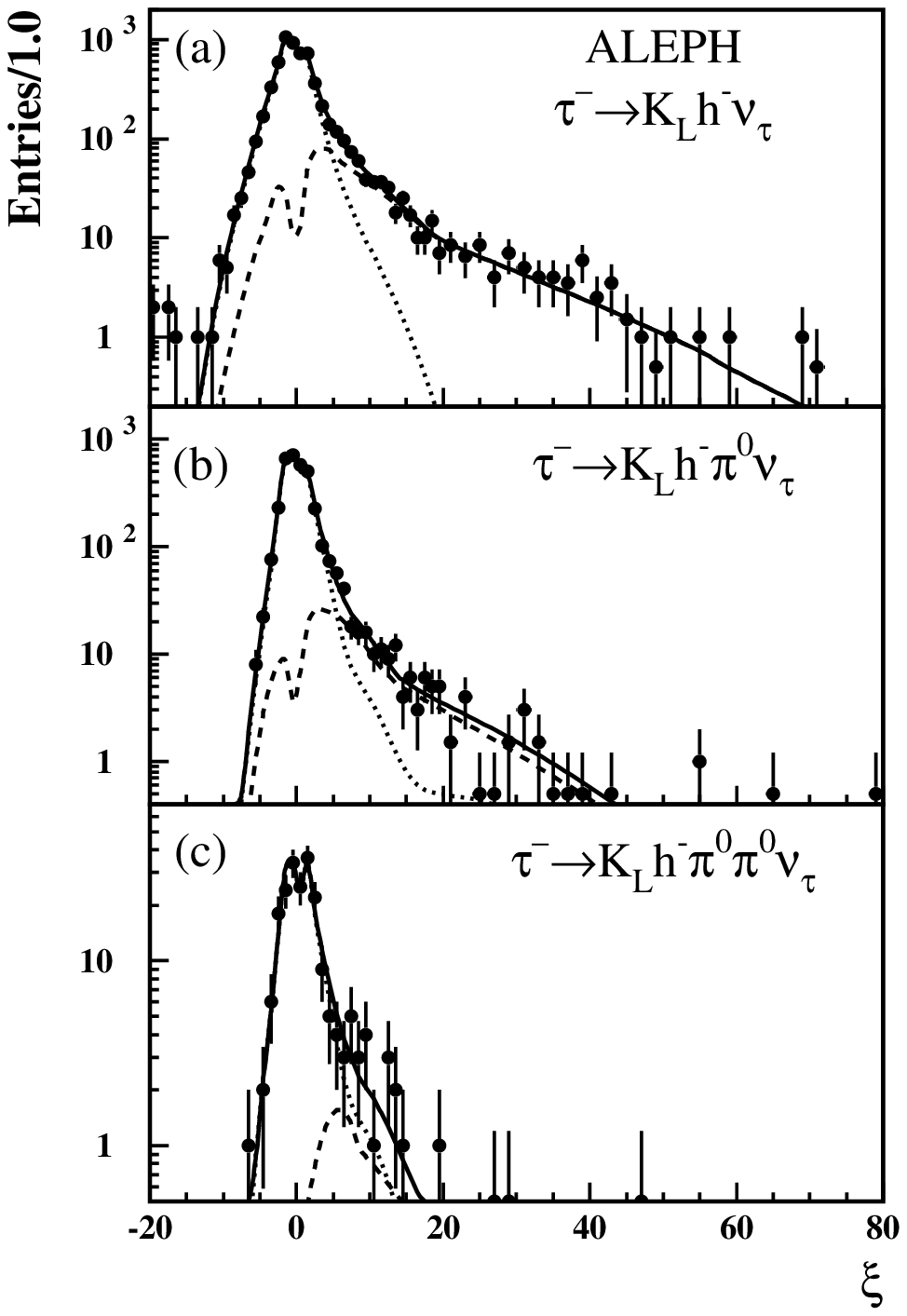}
\end{center}
\caption{\it The $\xi$ distributions for the $K^0_L$ sample (a)
in $h^-$ mode; (b) in $h^-\pi^0$ mode; 
(c) in $h^-\pi^0\pi^0$ mode. Data are given with error bars.
For $\xi>20$, the bin width is increased to 2.0.
The fits, signals ($h^-=\pi^-$) and $\tau$ backgrounds are shown as solid, 
dashed and dotted curves, respectively.
Even though the distributions for positive and negative $\xi$ are 
disconnected and vanish at $\xi=0$ by construction apart from binning effects, 
it is convenient to display the overall distribution in the same histogram.}
\label{xiklnpi}
\end{figure}

In the fits to the $\xi$ distributions, the 
means and the widths of all the Gaussian functions for the non-$K^0_L$ 
modes are left free in order to check the simulation and the calibration.
The channels involving a $K^-K^0_L$ pair are also taken into account by
incorporating the shapes from the simulation and fixing the fractions from 
the independent measurements of the branching ratios given in 
Table~\ref{exclk}.
The channels involving two neutral kaons cannot be 
separated from those involving a single neutral kaon because the
two $K^0_L$ showers overlap in the calorimeters.
A small difference is indeed observed in the simulated $\xi$ shapes,
but the fit is not sensitive to this effect. 
Table~\ref{exclkl} gives the results 
obtained in the $\xi$ fits. Figure~\ref{xiklnpi} shows all 
the $\xi$ distributions and the corresponding fits to the three 
$K^0_L$-enriched samples.  
\begin{table}
\begin{center}
\small
\begin{tabular}{|c|c|c|c|}\hline\hline
Mode & $K^0_L\pi^-$ & $K^0_L\pi^-\pi^0$ &$K^0_L\pi^-\pi^0\pi^0$ \\[3pt] 
\hline\rule{0pt}{13pt}
$\bar{\xi}~($non-$K^0_L$)
&$-0.04\pm0.03$&$0.00\pm0.03$&$-0.03\pm0.16$\\[3pt] \hline\rule{0pt}{13pt}
$\bar{\sigma_{\xi}}~($non-$K^0_L$)
&$0.98\pm0.02$&$0.96\pm0.02$&$1.00\pm0.08$\\[3pt] \hline
$f_{K^0_L}$ ($\%$) 
&$15.24\pm0.68$&$ 8.67\pm0.76$&$ 8.12\pm3.41$\\[3pt] \hline
$N_{K^0_L}$
&$937\pm41$&$299\pm26$&$17\pm7$\\[3pt] \hline
$\chi^2$/ndf
&72.5/69&46.0/41&14.0/21\\ \hline\rule{0pt}{13pt}
$N(K^0\overline{K^0}\pi^-\pi^0)$
&$2\pm1$&$16\pm6$&$6\pm2$\\ \hline
$N_{q\bar{q}}$
&$25\pm4$&$12\pm3$&$<1$\\ \hline
$B$ ($10^{-3}$)
&$9.28\pm0.45$&$3.47\pm0.53$&$<0.66~~(95\%\mbox{C.L.})$\\ \hline\hline
\end{tabular}
\caption{\it Results from the fits to $\xi$ distributions. The $\xi$
parameters for the non-$K^0_L$ modes, the fraction of $K^0_L$, 
the corresponding number of $K^0_L$'s, the test of goodness-of-fit in each
mode, the estimated backgrounds from $K^0\overline{K^0}\pi^-\pi^0$
and $q\bar{q}$, and the branching ratios obtained later.} 
\label{exclkl}
\end{center}
\end{table}   

\subsection{Branching ratios for decays with one $K^0$ only}

The relevant branching ratios are obtained from the 
measured $K^0_L$ yields using the efficiency matrix given
in Table~\ref{effkl}. Since the $K^0\overline{K^0}\pi^-(\pi^0)\nu_{\tau}$ 
channels can feed into the channels with a single neutral kaon, 
the subtraction of these backgrounds is necessary. 
The contribution of the $K^0_LK^0_L\pi^-$ mode is computed
using the measurement of the $K^0_SK^0_S\pi^-\nu_{\tau}$ channel, 
assuming $CP$ invariance. The latest 
measurements~\cite{k0decay,k0cleo} of the $K^0_SK^0_S\pi^-$ mode give an 
average $(0.24\pm0.05)\times 10^{-3}$, significantly 
smaller than the value $(0.75\pm0.38)\times 10^{-3}$~\cite{k0l3} 
used in the previous ALEPH analyses~\cite{1prong,hadbrs}, 
which relied on the assumption $B(\tau^-\to K^0_LK^0_L\pi^-\nu_{\tau})
/B(\tau^-\to K^0\overline{K^0}\pi^-\nu_{\tau})=1/4$.
After correcting for this effect, the published  
measurements~\cite{1prong,hadbrs} are in good agreement with the value
obtained in the present analysis.  
Since the $K^0_SK^0_L\pi^-$ mode can also give a contribution 
to all the final states as shown in Table~\ref{effk}, the subtraction
is performed using the branching ratio 
$B(\tau^-\to K^0_SK^0_L\pi^-\nu_{\tau})=(1.01\pm0.26)\times10^{-3}$ 
given by ALEPH~\cite{k0decay}. As for the decay mode 
$K^0\overline{K^0}\pi^-\pi^0$,
a first measurement of the branching ratio of $(0.31\pm0.12)\times 10^{-3}$ 
for the $K^0_SK^0_L\pi^-\pi^0$ part has been obtained by ALEPH~\cite{k0decay}.
For this mode, the ratios $K^0_SK^0_S:K^0_LK^0_L:K^0_LK^0_S$ of
1/4~:~1/4~:~1/2 are assumed, yielding the relevant contribution to
each final state given in Table~\ref{exclkl}.
Finally, the branching ratios 
are computed using the efficiencies given in Table~\ref{effkl}
and the numbers of neutral kaons, after doing the 
$K^0\overline{K^0}\pi^-\pi^0$ and $q\bar{q}$ background subtraction,
where the branching ratios for $K^0$ 
are given by assuming equal $K^0_S$ and $K^0_L$ contributions.
The results are given in Table~\ref{exclkl}.
\begin{table}
\begin{center}
\begin{tabular}{|c|c|c|c|c|c|}\hline\hline
Mode & $K^0_L\pi^-$ & $K^0_L\pi^-\pi^0$ & $K^0_L\pi^-\pi^0\pi^0$ 
& $K^0_LK^0_L\pi^-$ & $K^0_SK^0_L\pi^-$\\ \hline
$K^0_L\pi^-$        & 45.03& 4.20 & 0.23&32.11&3.86\\ \hline
$K^0_L\pi^-\pi^0$   &  2.34&24.42 & 1.17& 8.71&2.93\\ \hline
$K^0_L\pi^-\pi^0\pi^0$&0.24& 1.16 & 8.64& 0.53&3.46\\ \hline\hline
\end{tabular}
\caption{\it Efficiency matrix (in percent) 
for the exclusive $\tau$ decay modes 
involving one neutral kaon only. The feedthrough backgrounds from
the channels involving two neutral kaons are taken into account,
as detailed in the text.
The generated decay modes are given in the 
first row, and the reconstructed mode in the first column.}
\label{effkl}
\end{center}
\end{table}   

\section{Systematic uncertainties}

The sources of systematic uncertainties in this analysis are 
divided into several parts: general selection criteria, dE/dx response,
calorimetric measurements, feedthrough background, Monte Carlo
statistics and decay dynamics. The estimates are given in Table~\ref{sys}
and are detailed below.
\begin{table}
\begin{center}
\small
\begin{tabular}{|c|c|c|c|c|c|c|c|}\hline\hline
Source & Sel & dE/dx & Calori &
Bkg & MC stat & Dynam & Total \\ \hline
$K^-X$         & 0.3& 2.6& - & 0.7&0.4&  0 & 2.7\\ \hline
$K^-$          & 0.4& 1.6& - & 0.9&0.6&  0 & 2.0\\ \hline  
$K^-\pi^0$     & 0.7& 4.8& - & 2.0&1.2&  0 & 5.4\\ \hline
$K^-2\pi^0$    & 4.1&12.1& - &23.5&2.3& 5.8&27.5\\ \hline
$K^-3\pi^0$    &15.8& 4.1& - &22.1&4.8&10.0&29.6\\ \hline
$K^-K^0$       & 0.6& 1.4&3.2& 1.8&2.7& 5.0& 6.9\\ \hline
$K^-K^0\pi^0$  & 0.9&10.7&4.2& 6.9&4.2& 4.0&14.6\\ \hline
$\overline{K^0}\pi^-$
               & 0.6& -  &2.5& 2.2&1.4&  0 & 3.7\\ \hline
$\overline{K^0}\pi^-\pi^0$
               & 1.1& -  &5.4& 6.1&3.4& 6.0&10.8\\ \hline\hline  
\end{tabular}
\caption{\it Summary of systematic uncertainties (relative in percent).  
See text for details.}
\label{sys}
\end{center}
\end{table}

\subsection{Selection efficiency}

In this analysis, one-prong $\tau$ decays are classified into
$h^-$, $h^-\pi^0$, $h^-\pi^0\pi^0$ and $h^-\pi^0\pi^0\pi^0$ samples. The
uncertainties on the selection efficiencies, studied in 
Ref.~\cite{hadbrs}, correspond to the general $\tau$-pair selection, the 
lepton/hadron separation, the $\gamma/\pi^0$ recognition, the handling of
fake photons and the hadronic interactions. In the general $\tau$-pair 
selection, a $0.2\%$ uncertainty is given for all non-$K^0$ channels, 
while a $0.3\%$ uncertainty is assigned to the channels involving $K^0$. 
An uncertainty of $0.2\%$ on the lepton/hadron separation is 
estimated for all relevant channels, $i.e.$, without involving $\pi^0$'s.
The systematic uncertainty related to $\pi^0$ reconstruction increases
with the number of $\pi^0$'s involved. Errors of $0.6\%$, $2.6\%$ and $15.7\%$ 
are estimated for the channels involving one, two and three $\pi^0$'s,
respectively. Concerning fake photons,
a detailed investigation was presented in Ref.~\cite{hadbrs}.
It was found that there are some discrepancies between data and Monte
Carlo, resulting in corrections which are included in 
the efficiency matrix in Table~\ref{effk} and 
Table~\ref{effkl}: $(-1.49\pm0.32)\%$ 
for $K^-\nu_{\tau}$, $(-0.58\pm0.40)\%$ for $K^-\pi^0\nu_{\tau}$,  
$(+4.28\pm3.17)\%$ for $K^-\pi^0\pi^0\nu_{\tau}$,
$(+0.21\pm1.49)\%$ for $K^-\pi^0\pi^0\pi^0\nu_{\tau}$,
$(-1.62\pm0.54)\%$ for $h^-K^0\nu_{\tau}$,
$(+1.06\pm0.93)\%$ for $\overline{K^0}\pi^-\pi^0\nu_{\tau}$ and
$(-0.59\pm0.61)\%$ for $K^-K^0\pi^0\nu_{\tau}$.
The errors are taken as the systematic uncertainties. All the above
four contributions to the general selection criteria are put together and
given in Table~\ref{sys}.

\subsection{dE/dx measurement}

Three systematic effects are relevant to the dE/dx measurement.
Following Ref.~\cite{3prong}, the first part
is the pion calibration. The corresponding uncertainty is estimated
from the study of the inclusive mode, in which the fits to the $x_{\pi}$ 
distributions give two values for the error on the kaon fractions
(see Table~\ref{inclk1}), the first obtained from the 
three-parameter fits by leaving the $x_{\pi}(\pi)$ parameters free and
the second from the one-parameter fit when
fixing the $x_{\pi}(\pi)$ parameters. 
The pion calibration uncertainty is obtained from their 
quadratic difference, 
giving a $0.07\%$ uncertainty on the absolute kaon fraction. This value
is applied in a relative way to all channels involving charged kaons, 
yielding $2.3\%$ ($K^-X$), $1.4\%$ ($K^-$), $4.3\%$ ($K^-\pi^0$), 
$11.1\%$ ($K^-\pi^0\pi^0$), $3.2\%$ ($K^-\pi^0\pi^0\pi^0$), 
$0.9\%$ ($K^-K^0$) and $10.2\%$ ($K^-K^0\pi^0$), respectively. 

The second part is related to the shape of the 
pion resolution function $F_{\pi}(x_{\pi})$
derived from the muon samples. 
The largest contribution to the uncertainty comes from
the statistical errors in the parametrization of $F_{\pi}(x_{\pi})$, 
which correspond to relative uncertainties of: 
$0.7\%$ ($K^-X$), $0.4\%$ ($K^-$), $1.2\%$ ($K^-\pi^0$), 
$3.2\%$ ($K^-\pi^0\pi^0$), $0.9\%$ ($K^-\pi^0\pi^0\pi^0$), 
$0.3\%$ ($K^-K^0$) and $2.9\%$ ($K^-K^0\pi^0$), respectively. 
Another contribution arises from
the energy dependence of the parametrization using 
$\tau^-\to\mu^-\bar{\nu}_\mu\nu_\tau$ and $\gamma\gamma\to\mu^+\mu^-$
events. The corresponding uncertainties are: $0.1\%$ ($K^-X$), 
$0.1\%$ ($K^-$), $0.8\%$ ($K^-\pi^0$), 
$1.7\%$ ($K^-\pi^0\pi^0$) and $0.3\%$ ($K^-K^0$), respectively. 
The other modes with charged kaons are selected with
a 5 GeV/$c$ minimum momentum cut, thus reducing  
the effect studied above to a negligible level.

The third part comes from the $K/\pi$ separation,
which follows the calibrated velocity dependence obtained from pions.
There is no additional calibration error
for kaons in principle. However, for convenience the kaon resolution function,
Eq.~(\ref{fpi}), is constructed from Monte Carlo tracks, hence generating 
small statistical uncertainties of $1.0\%$ ($K^-X$), $0.8\%$ ($K^-$), 
$1.4\%$ ($K^-\pi^0$), $3.4\%$ ($K^-\pi^0\pi^0$), 
$2.4\%$ ($K^-\pi^0\pi^0\pi^0$), $1.0\%$ ($K^-K^0$) 
and $1.4\%$ ($K^-K^0\pi^0$), respectively.
These errors are translated into the systematic uncertainties for
each channel, and finally added to the other two dE/dx uncertainties 
in quadrature as shown in Table~\ref{sys}.

\subsection{Calorimetric measurements}

The systematic errors due to the uncertainties in the calorimetric
measurements are investigated by comparing the $\delta\phi$ and 
the $\delta_E$ distributions in data and Monte Carlo simulation.
From the statistical errors in the calibration procedure 
discussed in Section~5.3, an uncertainty is derived, through 
the $\delta\phi$ cuts, on the efficiencies    
for the $K^-K^0$ and $K^-K^0\pi^-$ modes, and the feedthrough
background from other $\tau$ decays, yielding 
systematic uncertainties of $1.0\%$ and $1.1\%$
the $K^-K^0$ and $K^-K^0\pi^-$ modes, respectively. 
Negligible uncertainties affect the other modes.
 
The precision of the $\delta_E$ measurement gives 
corresponding uncertainties of $2.9\%$ ($K^-K^0$), 
$3.9\%$ ($K^-K^0\pi^0$), $0.2\%$ ($K^-$) and 
$0.2\%$ ($K^-\pi^0$).

In selecting the $K^0_L$-enriched sample, a cut on the ratio 
of energy deposited in the ECAL and the HCAL 
is used to remove the background from $h^-\pi^0$
and $h^-\pi^0\pi^0$. To investigate the possible biases, two $\tau$
decay samples $h^-$ and $3h^-$ are studied. In the $h^-$ mode, events 
are required to have a positive $\delta\phi$ and a minimum
5 GeV HCAL energy. The comparison between data and Monte Carlo
gives a $(-2.7\pm1.0)\%$ correction to the efficiency estimated from
the Monte Carlo simulation. In the $3h^-$ mode, a similar study is performed, 
yielding a $(-0.2\pm0.7)\%$ correction. In this analysis, 
two hadrons are produced (charged hadron $+$ $K^0_L$) 
and an average of these two results yields
$(-1.5\pm1.0)\%$ for the $K^0h^-(\pi^0)$ modes, leading to a correction
included in the efficiency matrix in Table~\ref{effkl}. 

In obtaining the $\xi$ shape for the $K^0_L$ signal, an uncertainty
may arise from the validity of the calibration constants obtained from 
the non-$K^0_L$ samples when applying to the $K^0_L$ sample with
overlapping showers. To evaluate
this effect, the $3h$ $\tau$ decay mode is used to mimic the $K^0_L$ 
signal: $\delta_E$ and $\delta\phi$ are computed, using only one 
charged track momentum but the full calorimetric measurement for the 
three charged hadrons. As a result, the $\xi$ distribution looks 
very similar to the $K^0_L$ signal, providing
a relevant comparison between data and Monte Carlo simulation.
Applying the derived correction to the $\xi$ fit yields a bias
$\Delta f_{K^0_L}=(0.06\pm0.16)\%$ for the $K^0_L$ fraction 
in the $\overline{K^0}\pi^-$ mode.
A relative uncertainty on the corresponding branching ratio is therefore
estimated to be $1.6\%$. The same correction is also applied to the 
$\overline{K^0}\pi^-\pi^0$ mode, giving a $3.9\%$ relative uncertainty
on the branching ratio. 

Finally, the systematic effect due to the calorimetric calibration 
is investigated. Analogous to the systematic uncertainty study for 
dE/dx measurement, this effect is studied by fitting 
the mean and the resolution
of $\xi$ ($i.e.$, three-parameter vs one-parameter fits), giving $1.6\%$ 
and $3.4\%$ relative uncertainties on the branching ratios for
$\overline{K^0}\pi^-\nu_{\tau}$ and 
$\overline{K^0}\pi^-\pi^0\nu_{\tau}$, respectively.

\subsection{Background subtraction}

Since all one-prong $\tau$ decay with kaons are studied
simultaneously, except for $K^-\eta$ and the modes with two neutral kaons,
the uncertainties due to the feed-across effects are included in 
the branching ratios, taking into account the 
errors from the fits and the errors on the branching ratios used.

Background from $\rho$ and $a_1$ can feed into the $\overline{K^0}\pi^-$ and 
$\overline{K^0}\pi^-\pi^0$ modes when the $\pi^0$ leaks through an ECAL crack.
The cut on $\phi_{module}$ is designed to remove this background (see 
Section 7.1). However, the Monte Carlo simulation produces a somewhat 
broader shower shape in the calorimeters, leading to an uncertainty 
in calculating the barycentre in HCAL. To investigate this effect, a 
special sample is selected by requiring $\phi_{module}$ to be in a 
$5^\circ$ window centred on an ECAL crack and
$\xi>5$, enriching the background from $\rho$ and $a_1$ up to $50\%$ 
of the total. The comparison between data and Monte Carlo shows that
this background is over-estimated by a factor of $(1.2\pm0.1)$ in
the signal region. This correction is taken into account in obtaining
the $\xi$ shapes used in Section~7.2 for the $h^-\pi^0$ and $h^-\pi^0\pi^0$ 
modes with unreconstructed $\pi^0$'s.
The corresponding errors are treated as systematic
uncertainties, yielding $0.7\%$ for $\overline{K^0}\pi^-\nu_{\tau}$ and $1.0\%$
for $\overline{K^0}\pi^-\pi^0\nu_{\tau}$. 

Turning to the non-$\tau$ background, mainly $q\bar{q}$, the uncertainty 
arises from the dynamics to produce kaons. To investigate this effect, 
tighter cuts are applied 
on the hemisphere opposite to the one studied, rejecting
hemispheres with a hadronic invariant mass greater than 3.0 GeV/$c^2$.
More than $50\%$ of the $q\bar{q}$ background is removed at a cost of about 
$1.2\%$ in efficiency. Applying these cuts to the data reduces the fitted 
charged kaon yield by $85\pm15$. The Monte Carlo predicts that the kaons 
from $q\bar{q}$ background are reduced by $40\pm6$, while those from
$\tau$ decay decrease by $47\pm3$, resulting in a loss of
of $87\pm7$ kaons which agrees well with data 
and indicates that kaon production in these very low multiplicity $q\bar{q}$
events is well simulated within a $40\%$ uncertainty.
The same uncertainty is also given to the channels with 
neutral kaons.

\subsection{Monte Carlo statistics and dynamics}

Statistical uncertainties in the determination of the efficiency
matrices listed in Table~\ref{effk} and Table~\ref{effkl} are taken into
account. However, the most important source of uncertainty originates
from the decay models used in the generator, affecting
both the kaon and the $\pi^0$ momentum spectra. This effect may 
produce biases when determining the momentum-dependent $K/\pi$ separation, 
the $\pi^0$ reconstruction efficiency and the $K^0_L$ identification 
efficiency. All of these contributions are discussed below.

For the $K^-$, $K^-\pi^0$ and $\overline{K^0}\pi^-$ modes, the decay 
dynamics is well understood, and the corresponding uncertainties
are neglected. Since the inclusive $K^-X$ mode is dominated by the 
well established $K^-$ and $K^-\pi^0$ modes, and the fits to the 
$x_\pi$ distributions are performed in momentum bins, the uncertainty 
due to the unknown dynamics is neglected.
In the $K^-K^0$ and $K^-\pi^0\pi^0$ modes, 
the fit to the $x_{\pi}$ distribution is performed in momentum bins, 
reducing the uncertainty on the determination of $K/\pi$ separation
to a negligible level. 

In order to estimate the effect of $K^0_L$ and $\pi^0$ momentum spectra, 
the efficiencies are studied as a function of the hadronic mass. 
A mass uncertainty of $100$ MeV/$c^2$ is assigned for the modelling of the
$K^-K^0$, $K^-K^0\pi^0$, $K^-\pi^0\pi^0$ and $\overline{K^0}\pi^-\pi^0$ 
decay modes. The variations in the efficiencies are taken as  
systematic uncertainties, yielding $5.0\%$, $4.0\%$, $5.8\%$ and 
$6.0\%$ for the above decay modes, respectively. For the $K^-\pi^0\pi^0\pi^0$ 
mode, a $10\%$ uncertainty is assigned for the unknown dynamics.

\section{Investigation of mass spectra}

Mass spectra are investigated in the $K^-\pi^0$, $K^-\pi^0\pi^0$,
$K^-\pi^0\pi^0\pi^0$, $\overline{K^0}\pi^-$, $\overline{K^0}\pi^-\pi^0$ 
and $K^0K^-\pi^0$ modes. Because of the overlap between the  
hadronic showers, the $K^-K^0$ mode cannot be 
studied. The overlapping problem also affects the investigation of the total
invariant mass for $\overline{K^0}\pi^-\pi^0$ and  $K^0K^-\pi^0$.
Nevertheless, the study of subsystems such as $\pi^-\pi^0$ and $K^-\pi^0$
can still provide some useful information about the relevant dynamics.
A more complete assessment of the resonance structure taking into 
account all isospin-related channels is available~\cite{kms}.
   
\subsection{Mass spectra in the decays involving one $K^-$ only}

The study of mass spectra for the $K^-\pi^0$, $K^-\pi^0\pi^0$ and 
$K^-\pi^0\pi^0\pi^0$ decay modes is 
different from the conventional method, in which  
cuts are applied to isolate the charged kaon at a cost in efficiency.
Instead, this analysis generates the mass plots by fitting the number of 
charged kaons from the $x_{\pi}$ distribution in each mass slice. 

\begin{figure}
\begin{center}
\vspace{-2.0cm}
\epsfxsize 12cm
\epsffile{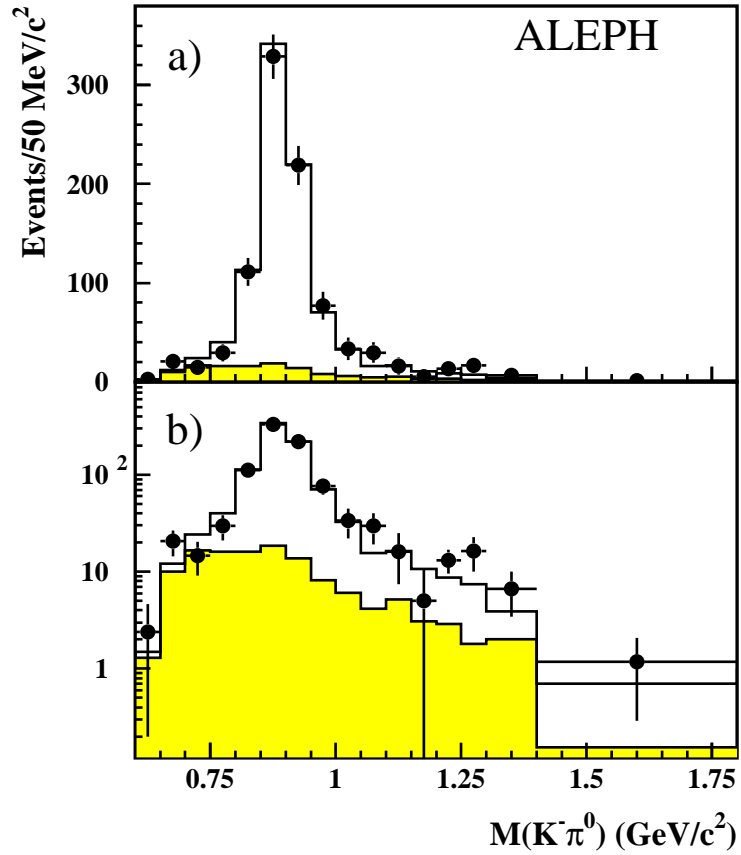}
\end{center}
\caption{\it The $K^-\pi^0$ invariant mass spectrum for the decay channel
$\tau^-\to K^-\pi^0\nu_{\tau}$: a) linear scale, b) log scale.
Data are shown as dots with error
bars and the Monte Carlo predictions are shown in the open histogram together
with the background distribution in shaded.}
\label{mkpi0}
\end{figure}

Since the minimum kaon momentum in the decay 
$\tau^-\to K^-\pi^0\nu_{\tau}$ is about 3.5 GeV/$c$, a cut is set for
excluding the low momentum tracks in the following analysis. 
A one-parameter fit is performed to extract the number 
of charged kaons in each mass slice, giving the mass plot shown in 
Fig.~\ref{mkpi0}, with a $14\%$ background level from other
$\tau$ decays involving charged kaons and by construction 
no background from pions. 
Except for a small excess on the high mass tail in the data which is 
discussed elsewhere~\cite{kms},
the $K^*(892)^-$ signal is well predicted by the Monte Carlo simulation, 
with the background mainly distributed on the low mass side.

\begin{figure}
\begin{center}
\vspace{-2.0cm}
\epsfxsize 14cm
\epsffile{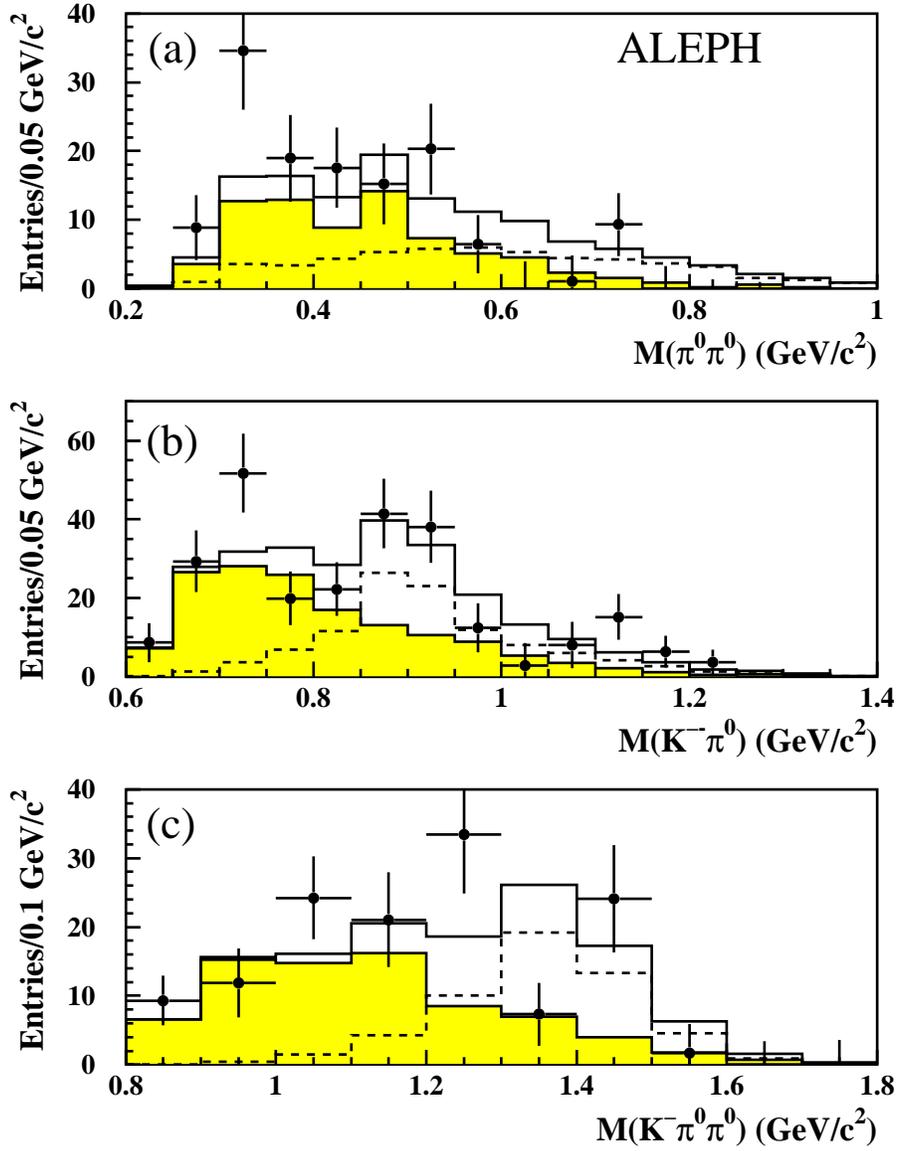}
\end{center}
\caption{\it The invariant mass spectra for the $K^-\pi^0\pi^0$ mode in data
(dots with error bars), Monte Carlo predictions (histograms) and the expected
background distribution (shaded histograms). The model predictions 
assuming $K_1(1400)$ dominance~\cite{was} are shown in dashed histograms.}
\label{mk2pi0}
\end{figure}

In the $K^-\pi^0\pi^0$ analysis, the requirement on the minimum kaon momentum
is raised to 5 GeV/$c$. The mass spectra for the $K^-\pi^0\pi^0$,
$K^-\pi^0$, and $\pi^0\pi^0$ systems are investigated accordingly.   
Figure~\ref{mk2pi0} shows these mass spectra for the $K^-\pi^0\pi^0$ mode. 
Many $\tau$ decay channels feed into this final state and contribute 
about $60\%$ of the total. 
However, since most backgrounds are
distributed in the region of low mass, some useful information related to the
channel $\tau^-\to K^-\pi^0\pi^0\nu_{\tau}$ can still be deduced.     
A peak at the $K^*(892)^-$ mass confirms the 
expected $K^*(892)^-\pi^0$ intermediate state. In general, $K_1(1270)$ and
$K_1(1400)$ can be produced in the $\overline{K}\pi\pi\nu_{\tau}$ channels. 
The model in Ref.~\cite{was} assumes $K_1(1400)$ dominance.
However, the $K^-\pi^0\pi^0$ mass spectrum in Fig.~\ref{mk2pi0} 
is consistent with an admixture of these two states. 

Because of low statistics in the $K^-\pi^0\pi^0\pi^0$ mode, 
the investigation of the mass spectra is difficult.
However the $\pi^0\pi^0\pi^0$ invariant mass can offer  
some useful information regarding the importance of 
the $K^-\eta$ channel. Only $5\pm4$ out of $22\pm7$ events are found in 
the $3\pi$ mass region (0.4--$0.65\,{\rm GeV}/c^2$), ruling out a 
dominant $\eta$ contribution, in agreement with measurements in the
$\eta\to\gamma\gamma$ mode~\cite{keta1}. 
Most of the charged kaons are found in the region
$M(3\pi^0)$ above 0.65 GeV/$c^2$, in which signal 
and background from $K^-K^0\pi^0\nu_{\tau}$
with $K^0_S\to\pi^0\pi^0$ overlap. 

\subsection{Mass spectra in the decays involving one $K^0_L$ only}

\begin{figure}
\begin{center}
\vspace{-2.0cm}
\epsfxsize 10cm
\epsffile{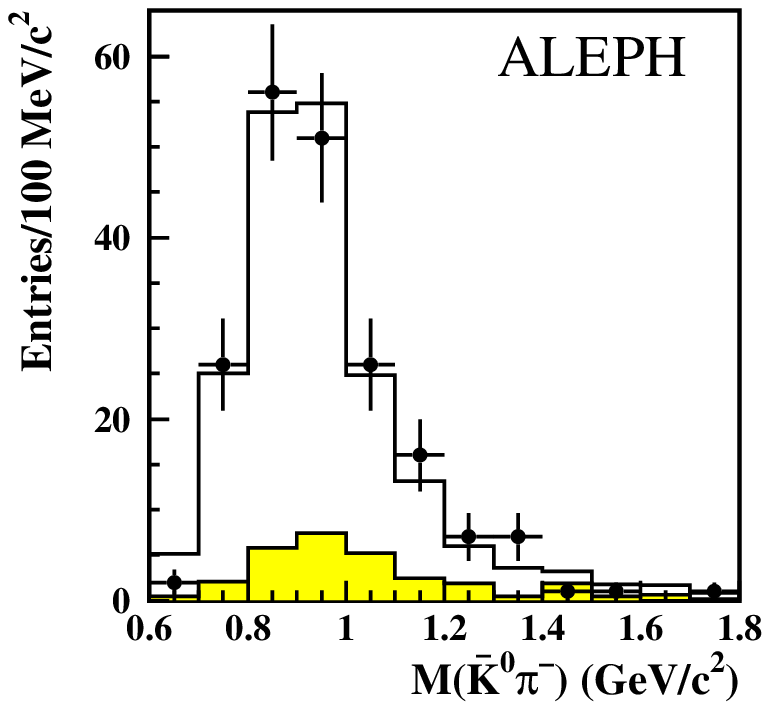}
\end{center}
\vspace{-0.5cm}
\caption{\it The $K^0_L\pi^-$ invariant mass spectrum for the decay channel
$\tau^-\to K^0_L\pi^-\nu_{\tau}$. Data are shown as dots with error
bars. Monte Carlo predictions and the expected background distribution
are shown in the open histogram and in the shaded one, respectively.}
\label{mk0pi}
\end{figure}

Candidates for the channels involving one $K^0_L$ are isolated by using 
cuts on both $\delta\phi$ and $\xi$. 
In the $K^0_L\pi^-$ mode, the charged pion momentum can be small, resulting
in a gap between the $K^0_L$ and $\pi^-$ showers.
The requirements $\delta\phi<-10$ and $\xi\geq 10$ remove almost all
non-$K^0_L$ background, yielding a purity of $86\%$ for the decay channel
$\tau^-\to K^0_L\pi^-\nu_{\tau}$. The $K^0_L$ flight direction is
defined from the colliding point to the barycentre of the energy 
deposited by the $K^0_L$ in the HCAL. The distribution of the invariant 
mass $M(K^0_L\pi^-)$ is shown in Fig.~\ref{mk0pi}, with a clear 
$K^*(892)^-$ signal seen.
The backgrounds from non-$K^*$ decay modes, such as the channels
$K^0_LK^0_L\pi^-\nu_{\tau}$ and $K^0_L\pi^-\pi^0\nu_{\tau}$, give a $10\%$
contamination, while the $q\bar{q}$ background contributes 
$4\%$. The Monte Carlo simulation is observed to be in good agreement 
with data, showing that the 
energy measured in the HCAL is adequately simulated. 

In the $K^0_L\pi^-\pi^0$ mode, the total invariant mass
provides less information because of the rather poorly reconstructed $K^0_L$
momentum and the low statistics. However, the study of 
the $\pi^-\pi^0$ invariant mass distribution can
give useful constraints on the relevant dynamics.

\begin{figure}
\begin{center}
\vspace{-2.0cm}
\epsfxsize 10cm
\epsffile{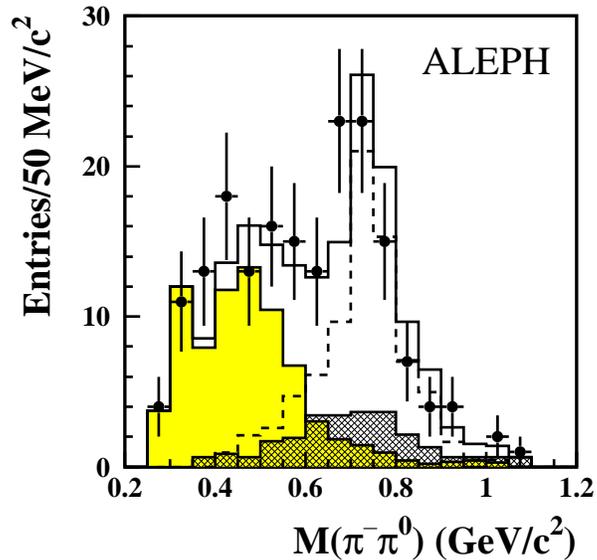}
\end{center}
\vspace{-0.5cm}
\caption{\it The $\pi^-\pi^0$ invariant mass spectrum in the decay 
channel $\tau^-\to~K^0_L\pi^-\pi^0\nu_\tau$. Data are shown as dots with error
bars. The Monte Carlo prediction (open histogram), the background 
contribution from $\rho^-/a^-_1$ (hatched histogram) and other $\tau$ 
backgrounds (shaded histogram) are shown, while the signal follows 
the dashed histogram.}
\label{m2pi1}
\end{figure}

The $K^0_L\pi^-\pi^0$ candidates are selected by
requiring the total measured energy in the HCAL to be greater than 
10 GeV, the angle $\alpha_{open}$ 
(defined in Section 6.2) less than $6^\circ$
and $\xi>5$. Under these additional requirements, 
the backgrounds from $\rho^-/a^-_1$, $K^*(892)^-$, 
$\overline{K^0}K^0\pi^-$ and $\overline{K^0}K^0\pi^-\pi^0$ are suppressed to
levels of $14\%$, $26\%$, $8\%$ and $8\%$, respectively. 
The $\pi^-\pi^0$ invariant mass is shown in 
Fig.~\ref{m2pi1}. An incoherent sum of a 
$\rho$ Breit-Wigner signal and the shape of the $\overline{K}^*\pi$ 
reflections is used to fit to the background-subtracted 
$\pi^-\pi^0$ invariant mass plot, giving $(72\pm12\pm10)\%$ for
the $\overline{K}\rho$ fraction, where the second $10\%$
uncertainty is due to the background subtraction. 
 
\subsection{Mass spectra in the decays involving a $K^-K^0_L$ pair}

\begin{figure}
\begin{center}
\vspace{-2.0cm}
\epsfxsize 10cm
\epsffile{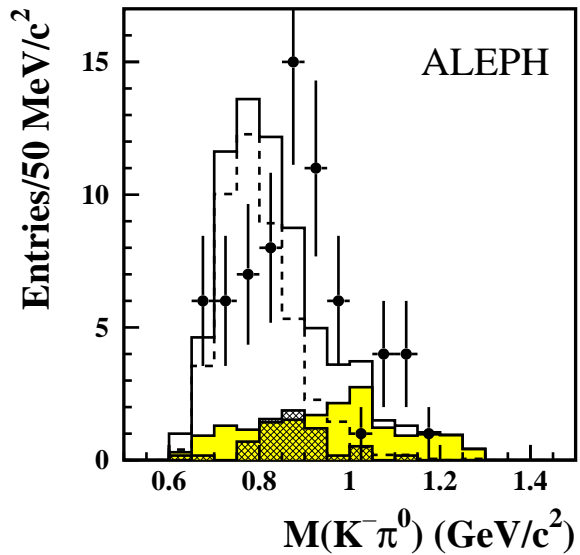}
\end{center}
\vspace{-0.5cm}
\caption{\it The $K^-\pi^0$ invariant mass spectrum for 
the $K^0_LK^-\pi^0$ candidates. Data (dots with error bars),
Monte Carlo prediction (open histogram), background contribution from
$K^-\pi^0(\pi^0)$ (hatched histogram) 
and other $\tau$ backgrounds (shaded histogram) are shown,
while the signal follows the dashed histogram.}
\label{mkk1}
\end{figure}

As mentioned before, the total invariant mass distributions 
in $K^-K^0_L\nu_{\tau}$ and $K^-K^0_L\pi^0\nu_{\tau}$ cannot be 
usefully studied. However, the $K^-\pi^0$ mass plot 
in $K^-K^0_L\pi^0\nu_{\tau}$ can be investigated. A requirement $x_{\pi}<-2$
is added for the $K^0_Lh^-\pi^0$ candidates, leading to a purity of
$65\%$. The $K^-\pi^0$ mass plot is shown in Fig.~\ref{mkk1} with a clear 
$K^*(892)^-$ signal observed. Although the feedthrough $K^-\pi^0(\pi^0)$ 
background can fake a $K^*(892)^-$ signal, it cannot account for 
such an excess in the $K^*(892)^-$ mass region. In the model of Ref.~\cite{was}
used in the Monte Carlo simulation, the $K^*K^0$ intermediate state in 
$K^-K^0\pi^0\nu_{\tau}$ is treated as strongly suppressed, which disagrees with
this observation. After subtraction of the background, 
the $K^-\pi^0$ invariant mass distribution for the decay 
$\tau^-\to K^0K^-\pi^0\nu_{\tau}$ is fitted with two components,
$K^*K$ and $\rho\pi$, showing that
the $K^*K$ intermediate state contribution in 
$K^0K^-\pi^0\nu_{\tau}$ is dominant (larger than $86\%$).
This observation is consistent with the result from the 
$K^+K^-\pi^-$ mode studied in Ref.~\cite{3prong}. 

\section{Results and discussion}

All measured branching ratios in this analysis are listed in 
Table~\ref{results}.
The inclusive branching ratio for $\tau^-\to K^-X\nu_{\tau}$ is
measured to be $(1.52\pm0.04\pm0.04)\%$, where $X$ represents any 
system of neutral particles.
The sum of all measured exclusive
branching ratios gives $(1.46\pm0.05\pm0.04)\%$, in 
agreement with the inclusive value.

The $(\overline{K}\pi)^-$ mode is studied in two final states, 
$K^-\pi^0$ and $K^0_L\pi^-$, both showing dominance of $K^*(892)^-$.  

The measurement of $\tau$ decay into $K^-\pi^0\pi^0\nu_\tau$ 
completes the study of the $(\overline{K}\pi\pi)^-$ final states 
started in Ref.~\cite{3prong,k0decay}.
The branching ratio $B(\tau^-\to K^-\pi^0\pi^0\nu_{\tau})
=(0.56\pm0.20\pm0.15)\times 10^{-3}$ is significantly smaller than
the value (1.1--$1.4) \times 10^{-3}$ predicted in 
the model of Ref.~\cite{fink}, but 
in good agreement with the value $0.4\times10^{-3}$ in Ref.~\cite{bal}. 
This channel suffers from a large background
contamination as evidenced by the study of invariant mass spectra.
Despite this, a dominant $K^*(892)^-\pi^0$ intermediate state is observed.
Background from $K^-\pi^0\pi^0\pi^0\nu_{\tau}$ is found to contribute to
the $K^-\pi^0\pi^0$ mode, resulting in a difference from the 
previous ALEPH measurement $B(\tau^-\to K^-\pi^0\pi^0\nu_{\tau})
=(0.80\pm0.20\pm0.20)\times 10^{-3}$~\cite{hadbrs}, wherein this background
was ignored.

Assuming that the decay 
$\tau^-\to \overline{K^0}\pi^-\pi^0\nu_{\tau}$
proceeds via incoherent intermediate states $\overline{K}\rho$ and 
$\overline{K}^*\pi$, one
can derive the following contributions:
\begin{eqnarray}
B(\tau^-\to(\overline{K}^*\pi)^-\nu_{\tau}
\to\overline{K^0}\pi^-\pi^0\nu_{\tau})
&=&(0.97\pm0.44\pm0.36)\times10^{-3},\\
B(\tau^-\to(\overline{K}\rho)^-\nu_{\tau}
\to\overline{K^0}\pi^-\pi^0\nu_{\tau})
&=&(2.50\pm0.57\pm0.44)\times10^{-3}.
\end{eqnarray}
These results are in agreement with the measurement based on the $K^0_S$
mode~\cite{k0decay}.

Finally, it is found that 
the decay $\tau^-\to K^-K^0\pi^0\nu_{\tau}$ mainly proceeds via 
the $K^*K$ intermediate state. 

\begin{table}
\begin{center}
\begin{tabular}{|l|c|}\hline\hline
Decay & $B~(10^{-3})$ \\ \hline
$\tau^-\to K^-X\nu_\tau$&
$15.20\pm0.40\pm0.41$\\ \hline
$\tau^-\to K^-\nu_\tau$&
$~6.96\pm0.25\pm0.14$\\ \hline
$\tau^-\to K^-\pi^0\nu_\tau$&
$~4.44\pm0.26\pm0.24$\\ \hline
$\tau^-\to K^-\pi^0\pi^0\nu_\tau$&
$~0.56\pm0.20\pm0.15$\\ \hline
$\tau^-\to K^-\pi^0\pi^0\pi^0\nu_\tau$ (excl. $\eta$)&
$~0.37\pm0.21\pm0.11$\\ \hline
$\tau^-\to K^-K^{0}\nu_\tau$&
$~1.62\pm0.21\pm0.11$\\ \hline
$\tau^-\to K^{-}K^{0}\pi^0\nu_\tau$&
$~1.43\pm0.25\pm0.21$\\ \hline
$\tau^-\to K^{-}K^{0}\pi^0\pi^0\nu_\tau$&
$<0.18~(95\%~\mbox{C.L.})$\\ \hline
$\tau^-\to \overline{K^0}\pi^{-}\nu_\tau$&
$~9.28\pm0.45\pm0.34$\\ \hline
$\tau^-\to \overline{K^0}\pi^{-}\pi^0\nu_\tau$&
$~3.47\pm0.53\pm0.37$\\ \hline
$\tau^-\to \overline{K^0}\pi^{-}\pi^0\pi^0\nu_\tau$&
$<0.66~(95\%~\mbox{C.L.})$\\ \hline\hline
\end{tabular}
\caption{Summary of branching ratios obtained in this analysis.
For the inclusive mode, $X$ represents any system of neutral particles.} 
\label{results}
\end{center}  
\end{table}

\section{Conclusion}

\begin{figure}
\begin{center}
\vspace{-2.0cm}
\epsfxsize 16cm
\epsffile{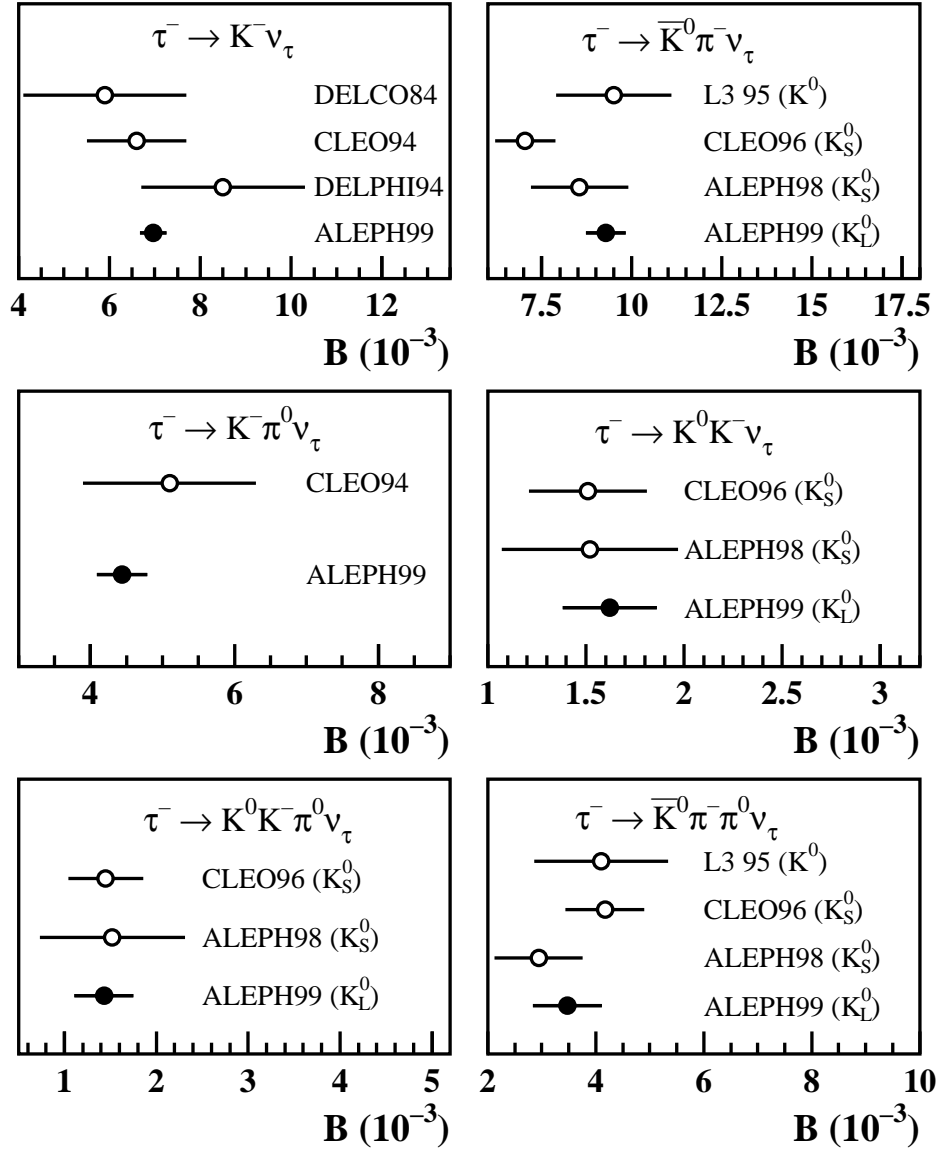}
\end{center}
\caption{\it The published branching ratios for one-prong $\tau$ decays
with kaons~\cite{pdg98}. 
The black dots correspond to this analysis. The detected 
$K^0$ modes are given in parentheses for each experiment.}
\label{allbrs}
\end{figure}

One-prong $\tau$ decays involving either charged and/or neutral kaons are
measured in this analysis. As already done in
the other two ALEPH analyses~\cite{3prong,k0decay}, the study of final
states is extended up to four hadrons involving kaons, giving a 
complete measurement for all sectors of $\tau$ decays involving kaons.
The results are given in Table~\ref{results} and also shown 
in Fig.~\ref{allbrs} to compare with other experiments~\cite{pdg98}. 
Agreement is observed with all the published data.
A forthcoming paper will combine all ALEPH
measurements on $\tau$ decays involving kaons and will give the relevant
physics implications~\cite{kms}.

\section*{Acknowledgements}

We wish to thank our colleagues in the CERN accelerator divisions for the
successful operation of the LEP storage ring. We also thank the engineers
and technicians in all our institutions for their support in
constructing and operating ALEPH. Those of us from nonmember states
thank CERN for its hospitality.

\end{document}